%% file: stochasticresolution.tex
\documentclass[11pt,oneside,a4paper]{article}
\usepackage[T1]{fontenc}
\usepackage[latin9]{inputenc}
\usepackage{verbatim}
\usepackage{float}
\usepackage{amsmath}
\usepackage{amssymb}
\usepackage{esint}
\usepackage{geometry}
\makeatletter

\floatstyle{ruled}
\newfloat{algorithm}{tbp}{loa}
\providecommand{\algorithmname}{Algorithm}
\floatname{algorithm}{\protect\algorithmname}

\usepackage{amssymb}
\usepackage{amsmath}
\usepackage{mathtools}
\usepackage{appendix}
\usepackage{color}
\usepackage{comment}
\usepackage{subfig} 
\usepackage{pstricks} 
\usepackage{float}
\usepackage{graphicx}
\usepackage{algorithmic}
\usepackage{amsthm}
\usepackage{stmaryrd}
\usepackage{authblk}

\usepackage{amsthm}
\usepackage{hyperref}               
\usepackage{graphicx}
\usepackage{color}              

\newtheoremstyle{break}
   {\topsep}{\topsep}%
   {\itshape}{}%
   {\bfseries}{}%
   {\newline}{}% 
\theoremstyle{break}

\newtheorem{theorem}{Theorem}%[section]
\newtheorem{lemma}[theorem]{Lemma}

\newtheorem{definition}{Definition}
\newtheorem{property}{Property}

\def\DEBUG{true}

\ifdefined\DEBUG
\definecolor{marekgreen}{RGB}{0,185,0}

  \def\rem#1{{\marginpar{\raggedright\scriptsize #1}}}
  \newcommand{\marr}[1]{\rem{\textcolor{marekgreen}{$\bullet$ #1}}}

  \def\rem#1{{\marginpar{\raggedright\scriptsize #1}}}
  \newcommand{\jusr}[1]{\rem{\textcolor{red}{$\bullet$ #1}}}

\else

  \newcommand{\marr}[1]{}

  \newcommand{\jusr}[1]{}

\fi

\makeatother

\begin{document}
\global\long\def\adj{\mbox{\footnotesize Adj}}

\global\long\def\chr#1{\mathbf{1}_{#1}}

\global\long\def\br#1{\left( #1 \right)}

\global\long\def\brq#1{\left[ #1 \right]}

\global\long\def\brw#1{\left\{  #1\right\}  }

\global\long\def\cut#1{\partial#1 }

\global\long\def\excond#1#2{\mathbb{E}\left[\left. #1 \right\vert #2 \right]}

\global\long\def\ex#1{\mathbb{E}\left[#1\right]}

\global\long\def\E{\mathbb{E}}

\global\long\def\exls#1#2{\mathbb{E}_{#1}\left[#2\right]}

\global\long\def\prcond#1#2{\mathbb{P}\left[\left. #1 \right\vert #2 \right]}

\global\long\def\setst#1#2{\left\{  \left.#1\right|#2\right\}  }

\global\long\def\set#1{\left\{  #1\right\}  }

\global\long\def\ind#1{\mathbf{1}\left[ #1 \right]}

\global\long\def\st#1{[#1] }

\global\long\def\size#1{\left|#1\right|}

\global\long\def\setstcol#1#2{\left\{  #1:#2\right\}  }

\global\long\def\set#1{\left\{  #1\right\}  }

\global\long\def\adj{\mbox{\footnotesize Adj}}

\global\long\def\indi#1{\chi\brq{#1}}

\global\long\def\evalat#1#2{ #1 \Big|_{#2}}

\global\long\def\Df#1#2{\frac{\partial#1}{\partial#2}}

\global\long\def\prls#1#2{\mathbb{P}_{#1}\left[ #2 \right]}
 \global\long\def\pr#1{\mathbb{P}\left[ #1 \right]}

\global\long\def\exls#1#2{\mathbb{E}_{#1}\left[ #2 \right]}

\global\long\def\excondls#1#2#3{\mathbb{E}_{#1}\brq{\left.#2\right|#3}}

\global\long\def\prcondls#1#2#3{\mathbb{P}_{#1}\brq{\left.#2\right|#3}}

\global\long\def\size#1{\left|#1\right|}

\global\long\def\setst#1#2{\left\{  #1\left|#2\right.\right\}  }

\global\long\def\setstX#1#2{\left\{  \left.#1\right|#2\right\}  }

\global\long\def\setstcol#1#2{\left\{  #1:#2\right\}  }

\global\long\def\set#1{\left\{  #1\right\}  }

\global\long\def\adj#1{\delta\br{#1}}

\global\long\def\indi#1{\chi\brq{#1}}

\global\long\def\evalat#1#2{ #1 \Big|_{#2}}

\global\long\def\Df#1#2{\frac{\partial#1}{\partial#2}}

\global\long\def\hx#1{\hat{x}_{#1}}

\global\long\def\eps{\varepsilon}

\global\long\def\M{{\cal M}}

\global\long\def\Min{{\cal M}}

\global\long\def\M{{\cal M}}

\global\long\def\Iin{{\cal I}^{in}}

\global\long\def\Iout{{\cal I}^{out}}

\global\long\def\kin{k^{in}}

\global\long\def\kout{k^{out}}

\global\long\def\M{{\cal M}}

\global\long\def\R{R\br x}

\global\long\def\indi#1{\chi\brq{#1}}

\global\long\def\chr#1{\mathbf{1}_{#1}}

\global\long\def\X#1{\hat{X}_{#1}}

\global\long\def\E{\hat{E}}

\global\long\def\block#1{\Gamma\br{#1}}

\global\long\def\zblock#1{\Gamma_{0}\br{#1}}

\global\long\def\outblock#1{\Gamma^{out}\br{#1}}

\global\long\def\inblock#1{\Gamma^{in}\br{#1}}

\global\long\def\jblock#1{\Gamma^{j}\br{#1}}

\global\long\def\Frac#1#2{#1\left/\br{#2}\right.}

\global\long\def\P#1{{\cal P}\br{#1}}

\global\long\def\I{{\cal I}}

\global\long\def\setphi#1#2#3{\phi\brq{#1,#2}\br{#3}}

\global\long\def\support#1{\mbox{supp}\br{#1}}

\global\long\def\act#1{act\br{#1}}

\global\long\def\brqbb#1{\llbracket#1\rrbracket}

\global\long\def\Sact{\act S}

\global\long\def\Ract{\act{\R}}

\global\long\def\marro#1{\mbox{\textcolor{red}{#1}}}

\global\long\def\B{{\cal B}}

\pagestyle{headings}
\author[1]{Marek Adamczyk \thanks{Supported by the ERC StG project PAAl no.~259515.}} 
\title{Non-negative submodular stochastic probing via stochastic contention resolution schemes}
\affil[1]{Department of Computer, Control, and Management Engineering, Sapienza University of Rome, Italy, \texttt{adamczyk@dis.uniroma1.it.}} \maketitle 
\begin{abstract}
In a \emph{stochastic probing }problem we are given a universe $E$,
where each element $e\in E$ is \emph{active} independently with probability
$p_{e}\in\brq{0,1}$, and only a \emph{probe} of $e$ can tell us
whether it is active or not. On this universe we execute a process
that one by one probes elements --- if a probed element is active,
then we have to include it in the solution, which we gradually construct.
Throughout the process we need to obey \emph{inner }constraints on
the set of elements taken into the solution, and \emph{outer} constraints
on the set of all probed elements. The objective is to maximize a
function of successfully probed elements.

This abstract model was presented by Gupta~and~Nagarajan~(IPCO'13),
and provides a unified view of a number of problems. Adamczyk, Sviridenko, Ward~(STACS'14)
gave better approximation for matroid environments and linear objectives.
At the same time this method was easily extendable to settings, where
the objective function was monotone submodular. However, the case
of non-negative submodular function could not be handled by previous
techniques.

In this paper we address this problem, and our results are twofold.
First, we adapt the notion of contention resolution schemes of Chekuri, Vondr\'{a}k, Zenklusen~(SICOMP'14) to show that we can optimize
non-negative submodular functions in this setting with a constant factor
loss with respect to the deterministic setting. Second, we show
a new contention resolution scheme for transversal matroids, which yields better approximations in the stochastic
probing setting than the previously known tools. The rounding procedure underlying the scheme
can be of independent interest --- Bansal, Gupta, Li, Mestre, Nagarajan, Rudra~(Algorithmica'12) gave 
two seemingly different algorithms for stochastic matching
and stochastic $k$-set packing problems with two different analyses, but
we show that our single technique can be used to analyze both their algorithms.
\end{abstract}
\newpage
\section{Introduction}

\iffalse
Uncertainty in input data is a common feature of most practical problems,
and research in finding good solutions (both experimental and theoretical)
for such problems has a long history dating back to 1950~\cite{Beale:convex,Dantzig:uncertainty}.\fi

Stochastic variants of optimization problems were considered already in 1950~\cite{Beale:convex,Dantzig:uncertainty}, but only in recent years a significant attention was brought to approximation algorithms for stochastic variants of combinatorial problems. In this paper we consider adaptive stochastic optimization problems in the framework
of Dean~et~al.~\cite{DBLP:journals/mor/DeanGV08} who presented a stochastic knapsack problem.
Since the work of Dean~et~al.\ a
number of problems in this framework were introduced~\cite{DBLP:conf/icalp/ChenIKMR09,DBLP:conf/latin/GoemansV06,DBLP:conf/stoc/GuhaM07,DBLP:conf/soda/GuhaM07,DBLP:conf/wine/AsadpourNS08,DBLP:conf/focs/GuptaKMR11,DBLP:conf/soda/DeanGV05}.
Gupta~and~Nagarajan~\cite{DBLP:conf/ipco/GuptaN13}
presented an abstract framework for a subclass of adaptive stochastic
problems giving a unified view for stochastic matching~\cite{DBLP:conf/icalp/ChenIKMR09}
and sequential posted pricing~\cite{DBLP:conf/stoc/ChawlaHMS10}.
Adamczyk et al.~\cite{DBLP:conf/stacs/AdamczykSW14} generalized
the framework by also considering monotone submodular functions in
the objective. In this paper we generalize the framework even further
by showing that also maximizing a non-negative submodular function
can be considered in the probing model. On the way we develop a randomized
procedure for transversal matroids which can be used to improve approximation
for the $k$-set packing problem~\cite{Bansal:woes}.

Below paper enhances the iterative randomized rounding for points from matroid polytopes that was presented in~\cite{DBLP:conf/stacs/AdamczykSW14}. The analysis from~\cite{DBLP:conf/stacs/AdamczykSW14} does not easily carry over when the objective submodular function is non-monotone. To handle non-monotone objectives 
we are making use of \emph{contention resolution schemes} introduced by~Chekuri~et~al.~\cite{DBLP:conf/stoc/VondrakCZ11}. Contention resolution schemes in the context of stochastic probing already were used by Gupta~and~Nagarajan~\cite{DBLP:conf/ipco/GuptaN13}. Recently, Feldman et al.~\cite{DBLP:journals/corr/FeldmanSZ15} presented online version of contention resolution schemes which, on top of applications for online settings, yield good approximations for stochastic probing problem for a broader set of constraints than before ---  most notably, for inner knapsack constraints and \emph{deadlines}.

Our paper fills the gaps between and merges results from basically four different paper~\cite{DBLP:conf/stacs/AdamczykSW14,DBLP:conf/stoc/VondrakCZ11,DBLP:conf/ipco/GuptaN13,DBLP:conf/focs/FeldmanNS11}. That is the reason why this paper comes with diverse contributions: we improve the bound on measured greedy algorithm of Feldman
et al.~\cite{DBLP:conf/focs/FeldmanNS11}; adjust contention resolution schemes to stochastic probing setting in a way in which submodular optimization is be possible; use iterative randomized rounding technique to develop contention resolution schemes; moreover, we revisit the algorithms of Bansal et al.~\cite{Bansal:woes}. 

Below we present the necessary background.

\subsubsection*{The probing model}

We describe the framework following~\cite{DBLP:conf/ipco/GuptaN13}.
We are given a universe $E$, where each element $e\in E$ is \emph{active
}with probability $p_{e}\in[0,1]$ independently. The only way to
find out if an element is active, is to \emph{probe }it. We call a
probe \emph{successful} if an element turns out to be active. On universe
$E$ we execute an algorithm that probes the elements one-by-one.
If an element is active, the algorithm is forced to add it to the
current solution. In this way, the algorithm gradually constructs
a solution consisting of active elements.

Here, we consider the case in which we are given constraints on both
the elements probed and the elements included in the solution. Formally,
suppose that we are given two independence systems of downward-closed
sets: an \emph{outer} independence system $\br{E,\Iout}$ restricting
the set of elements probed by the algorithm, and an \emph{inner} independence
system $\br{E,\Iin}$, restricting the set of elements taken by the
algorithm. We denote by $Q^{t}$ the set of elements probed in the
first $t$ steps of the algorithm, and by $S^{t}$ the subset of active
elements from $Q^{t}$. Then, $S^{t}$ is the partial solution constructed
by the first $t$ steps of the algorithm. We require that at each
time $t$, $Q^{t}\in\Iout$ and $S^{t}\in\Iin$. Thus, at each time
$t$, the element $e$ that we probe must satisfy both $Q^{t-1}\cup\{e\}\in\Iout$
and $S^{t-1}\cup\{e\}\in\Iin$. The goal is to maximize expected value
$\ex{f\br S}$ where $f:2^{E}\rightarrow\mathbb{R}_{\geq0}$ and $S$
is the set of all successfully probed elements.

We shall denote such a stochastic probing problem by $\br{E,p,\Iin,\Iout}$
with function $f$ stated on the side, if needed.

\subsubsection*{Submodular Optimization}

A set function $f:2^{E}\mapsto\mathbb{R}_{\ge0}$ is \emph{submodular},
if for any two subsets $S,T\subseteq E$ we have $f\br{S\cup T}+f\br{S\cap T}\leq f\br S+f\br T$.
Without loss of generality, we assume also that $f\br{\emptyset}=0$.

The \emph{multilinear extension} $F:[0,1]^{E}\mapsto\mathbb{R}_{\ge0}$
of $f$, whose value at a point $y\in\brq{0,1}^{E}$ is given by 
\[
F\br y=\sum_{A\subseteq E}f\br A\cdot\prod_{e\in A}y_{e}\prod_{e\not\in A}\br{1-y_{e}}.
\]
Note that $F\br{\chr A}=f\br A$ for any set $A\subseteq E$, so $F$
is an extension of $f$ from discrete domain $2^{E}$ into a real
domain $\brq{0,1}^{E}$. The value $F(y)$ can be interpreted as the
expected value of $f$ on a random subset $A\subseteq E$ that is
constructed by taking each element $e\in E$ with probability $y_{e}$.

\subsubsection*{Contention Resolution Schemes}

Consider a ground set of elements $E$ and an down-closed family ${\cal I}\subseteq2^{E}$
of $E$'s subsets --- we call $\br{E,{\cal I}}$ an \emph{independence
system.} Let ${\cal P}\br{{\cal I}}$ be the convex hull of characteristic
vectors of sets from ${\cal I}$. Given $x\in\P{\I}$ we define $R\br x$
to be a random set in which every element $e\in E$ is included in
$R\br x$ with probability $x_{e}$; set $\R$ defined like that is
used frequently throughout the paper.

Chekuri et al.~\cite{DBLP:conf/stoc/VondrakCZ11} presented a framework
of contention resolution schemes (CR schemes) that allows to maximize
non-negative submodular functions for various constraints. The following
definition and theorem come from~\cite{DBLP:conf/stoc/VondrakCZ11}.

\begin{definition} \label{def:crscheme}Let $\br{E,{\cal I}}$ be
independence system. For $b,c\in\brq{0,1}$, a $\br{b,c}$-balanced
CR scheme $\pi$ for ${\cal P}\br{{\cal I}}$ is a randomized procedure
that for every $x\in b\cdot{\cal P}\br{{\cal I}}$ and $A\subseteq E$,
returns a random\emph{ }set $\pi_{x}\br A$ such that: 
\begin{enumerate}
\item always $\pi_{x}\br A\subseteq A\cap\support x$ and $\pi_{x}\br A\in{\cal I}$,
\item $\pr{e\in\pi_{x}\br{A_{1}}}\geq\pr{e\in\pi_{x}\br{A_{2}}}$ whenever
$e\in A_{1}\subseteq A_{2}$,
\item \label{enu:crscheme-marginal}for all $e\in\support x$, $\prcond{e\in\pi_{x}\br{R\br x}}{e\in R\br x}\geq c$.
\end{enumerate}

\end{definition}

\begin{theorem} \label{thm:chekurimultilinearbound} Let $\br{E,\I}$
be an independence system. Let $f:2^{E}\rightarrow\mathbb{R}$ be
a non-negative submodular function with multilinear relaxation $F$,
and $x$ be a point in $b\cdot{\cal P}\br{{\cal I}}$. Let $\pi$ be a  $\br{b,c}$-balanced
CR scheme for ${\cal P}\br{{\cal I}}$, and let $S=\pi_{x}\br{R\br x}$.
If $f$ is monotone then $\ex{f\br S}\geq c\cdot F\br x.$ Furthermore,
there is a function $\eta_{f}:2^{E}\rightarrow2^{E}$ that depends
on $f$ and can be evaluated in linear time, such that even for $f$
non-monotone $\ex{f\br{\eta_{f}\br S}}\geq c\cdot F\br x.$

\end{theorem} Function $\eta_{f}\br S$ represents a \emph{pruning
}operation that removes from $S$ some elements. To prune a set $S$
with pruning function $\eta_{f}$, an arbitrary ordering of the elements
of $E$ is fixed: for simplicity of notation let $E=\set{1,...,\size E}$
which gives a natural ordering. Starting with $S^{prun}=\emptyset$
the final set $S^{prun}=\eta_{f}\br S$ is constructed by going through
all elements of $E$ in the given order. When considering an element
$e$, $S^{prun}$ is replaced by $S^{prun}+e$ if $f\br{S^{prun}+e}-f\br{S^{prun}}\geq0$.

Note that a pruning operation like that is not possible to execute
in the probing model since we commit to elements. We address this
issue in Section~\ref{sub:Stochastic-contention-resolution}, where we show how
to perform on-the-fly pruning. 

\subsubsection*{Stochastic $k$-set packing}

We are given $n$ elements/columns, where each element $e\in\brq n$
has a random profit $v_{e}\in\mathbb{R}_{+}$, and a random $d$-dimensional
size $S_{e}\in\{0,1\}^{d}$. The sizes are independent for different
elements, but $v_{e}$ can be correlated with $S_{e}$, and the coordinates
of $S_{e}$ also might be correlated between each other. Additionally,
for each element $e$, there is a set $C_{e}\subseteq[d]$ of at most
$k$ coordinates such that each size vector $S_{e}$ takes positive
values only in these coordinates, i.e., $S_{e}\subseteq C_{e}$ with
probability 1. We are also given a capacity vector $b\in\mathbb{Z}_{+}^{d}$
into which elements must be packed. We say that column $S_{e}$ outcomes
are monotone if for any possible realizations $a,b\in\set{0,1}^{d}$
of $S_{e}$, we have $a\leq b$ or $b\leq a$ coordinate-wise. A strategy
probes columns one by one, obeying the packing constraints, and its
goal is to maximize the expected outcome of taken columns. Bansal
et al.~\cite{Bansal:woes} gave a $2k$-approximation algorithm for
the problem, and a $\br{k+1}$-approximation algorithm when they assume
that the outcomes of size vectors $S_{e}$ are monotone.
The technique from~\cite{DBLP:conf/stacs/AdamczykSW14} allows to get a $(k+1)$-approximation
for the problem without the assumption of monotone column outcomes.

\subsection{Our contributions at a high level}

There are two main contributions of the paper. First is Theorem~\ref{thm:maintheorem}
which implies that non-negative submodular optimization in the probing
model is possible if we are given CR schemes. Our second contribution
is the improved insight for transversal matroids in the context of
stochastic probing, captured in Theorem~\ref{thm:transversal-matroids}
and new analyses of algorithms from~\cite{Bansal:woes}
for stochastic $k$-set packing and stochastic matching.
It is based on the iterative
randomized rounding technique of~\cite{DBLP:conf/stacs/AdamczykSW14}.

\subsubsection{Non-negative submodular optimization via CR schemes in the probing
model}

To obtain our results we extend the framework of CR schemes into the
probing model, and we define a \emph{stochastic contention resolution
scheme (stoch-CR scheme)}. Define a polytope 
\[
\P{\Iin,\Iout}=\setst x{x\in\P{\Iout},p\cdot x\in\P{\Iin},x\in\brq{0,1}^{E}}.
\]
By $\act S$ we denote the subset of active elements of set $S$.
Note that event $e\in\act{\R}$ means both that $e\in\R$ and that
$e$ is active; this event has probability $p_{e}x_{e}$.

\begin{definition} \label{def:stochCRscheme}
Let $\br{E,p,\Iin,\Iout}$ be a stochastic probing problem.
For $b,c\in\brq{0,1}$,
a $\br{b,c}$-balanced \emph{stoch-CR scheme }$\bar{\pi}$ for a polytope
$\P{\Iin,\Iout}$ is a probing strategy that for every $x\in b\cdot\P{\Iin,\Iout}$ and $A\subseteq E$, obeys outer constraints $\Iout$, and the returned random set $\bar{\pi}_{x}\br A$ satisfies the following:
\begin{enumerate}
\item $\bar{\pi}_{x}\br A$ consists only of active elements,
\item $\bar{\pi}_{x}\br A\subseteq A\cap\support x$ and $\bar{\pi}_{x}\br A\in\Iin$,
\item \label{item:stochCRschemeaverage}$\prcond{e\in\bar{\pi}_{x}\br{\R}}{e\in\act{\R}}\geq c$,
\item $\pr{e\in\bar{\pi}_{x}\br{A_{1}}}\geq\pr{e\in\bar{\pi}_{x}\br{A_{2}}}$
whenever $e\in A_{1}\subseteq A_{2}$.
\end{enumerate}
\end{definition}In Section~\ref{sec:Non-negative-submodular-optimization-via-stochasitc-CR-schemes}
we present a mathematical program that models our problem. Solving
the program, getting $x^{+},$ and running the stoch-CR scheme $\bar{\pi}_{x^{+}}$
on $R\br{x^{+}}$ constitute the algorithm from the below Theorem.

\begin{theorem}\label{thm:maintheorem} Consider a stochastic probing
problem $\br{E,p,\Iin,\Iout}$, where we need to maximize a non-negative
submodular function $f:2^{E}\mapsto\mathbb{R}_{\geq0}$. If there
exists a $\br{b,c}$-balanced stoch-CR scheme $\bar{\pi}$ for $\P{\Iin,\Iout}$,
then there exists a probing strategy whose expected outcome is at
least $c\br{b\cdot e^{-b}-o\br 1}\cdot\ex{f\br{OPT}}$. 

\end{theorem}

To put this Theorem into perspective. There exists a $\br{b,\br{\frac{1-e^{-b}}{b}}^{k}}$-balanced scheme for intersection of $k$ matroids~\cite{DBLP:conf/stoc/VondrakCZ11}, and there exists a $\br{b,(1-b)^k}$-balanced
ordered scheme for intersection of $k$ matroids~\cite{DBLP:conf/stoc/ChawlaHMS10,DBLP:journals/corr/FeldmanSZ15}. For example when $\kin=\kout=1$, Lemma~\ref{lem:combining} and the above Theorem, after plugging appropriate number $b$, yield
approximation of $0.13$. See Section~\ref{sub:Stochastic-probing-for-varous-constraints}
for discussion on other types of constraints.

\subsubsection{Optimization for transversal matroids}

\paragraph{Stochastic probing on intersection of transversal matroids}

We improve upon the 
\iffalse
$\frac{1}{4\br{\kin+\kout}}\cdot e^{-\frac{1}{2\br{\kin+\kout}}}$\fi
approximation described in the previous Section,
if we assume the constraints are intersections of transversal matroids.
We do it by developing a new stoch-CR scheme. This scheme is direct
in the sense that we do not construct it by applying first a scheme
for outer and then for inner constraints, as in Lemma~\ref{lem:combining}.

\begin{lemma}\label{lem:stochCRscheme}

There exists a $\br{b,\frac{1}{1+b\cdot\br{\kin+\kout}}}$-balanced
stoch-CR scheme when constraints are intersections of $\kin$ inner
and $\kout$ outer transversal matroids.

\end{lemma}
For $\kin=\kout=1$ the above Lemma together with Theorem~\ref{thm:maintheorem} give approximation factor of $0.15$, so a modest improvement over $0.13$,
but it gets significantly better for larger values of $\kin+\kout$.
With $k=\kin +\kout$ we plug $b=\frac{2}{\sqrt{1+4k}+1}$ and we use Theorem~\ref{thm:maintheorem}
to conclude the following Theorem.
\begin{theorem}\label{thm:transversal-matroids}
For maximizing non-negative submodular function in the probing model
with $\kin$ inner and $\kout$ outer transversal matroids there exists
an algorithm with approximation ratio of $$\frac{1}{k+\sqrt{k+\frac{1}{4}}+\frac{1}{2}}\br{1-\Theta\br{\frac{1}{\sqrt{k}}}}.$$ 
\end{theorem}
There are further applications of the techniques used
in Lemma~\ref{lem:stochCRscheme}.

\paragraph*{Regular CR scheme for transversal matroids}

When $\kin=0$ this scheme yields $\br{b,\frac{1}{1+b\cdot k}}$-balanced
CR scheme for deterministic setting. So far only a $\br{b,(1-b)^k}$-balanced
\emph{ordered} scheme and a $\br{b,\br{\br{1-e^{b}}/b}^{k}}$-balanced
scheme were known~\cite{DBLP:conf/stoc/VondrakCZ11}; see section~\ref{sub:Stochastic-probing-for-varous-constraints}
for the definition of ordered scheme. Our scheme can be seen as an
improvement when one looks at the $\max_{b}\br{b\cdot c}$ --- first
one yields $\frac{1}{e(k+1)}$, second $\frac{2}{e\br{k+1}}$(for
$k$ big\footnote{The precise value is smaller than $\frac{1}{k+1}$
starting $k\geq4$.}), and we get $\frac{1}{k+1}$. 

\paragraph{Stochastic $k$-set packing and stochastic matching}
Here we want to argue that the martingale-based analysis of the algorithm for transversal matroids
can be of independent interest. 
In the bipartite stochastic matching problem~\cite{Bansal:woes}
the inner constraints define bipartite matchings and outer constraints
define general $b$-matchings --- both are intersections of $2$ transversal
matroids. 
Stochastic $k$-set packing problem does not belong to our framework, but we have already defined it in previous subsection.
%Therefore, scheme from Lemma~\ref{lem:stochCRscheme} yields a 5-approximation for weighted case when $b=1$. 

Bansal et al.~\cite{Bansal:woes} presented an algorithm
for the $k$-set packing, and have proven that it yields a $\br{k+1}$-approximation but when assuming
that the column outcomes are monotone. They also presented a 3-approximation algorithm
for the stochastic matching problem.
Both algorithms first choose a subset of elements: for $k$-set packing the elements are chosen
independently, and for stochastic matching the elements are chosen using a dependent rounding
procedure of Gandhi et al.~\cite{DBLP:journals/jacm/GandhiKPS06}.
After that they probe elements in a random order.
Bansal et al.~\cite{Bansal:woes} gave two quite different analyses of the two algorithms,
while we show how both algorithms can be analyzed using our martingale-based technique
for transversal matroids.
First, we show that their algorithm for set packing is in fact a $(k+1)$-approximation even without the monotonicity assumption (see Appendix~\ref{sec:Stochastic--set-packing}).
Second, we also show that the our technique of analysis provides a proof of 3-approximation of their algorithm for stochastic matching (see Appendix~\ref{sec:Stochastic-Matching}).

\section{Optimization via stoch-CR schemes\label{sec:Non-negative-submodular-optimization-via-stochasitc-CR-schemes}}

\subsubsection*{Mathematical program}

Another extension of $f$ studied in \cite{Calinescu2011} is given
by: 
\[
f^{+}(y)=\max\setst{\sum_{A\subseteq E}\alpha_{A}f(A)}{\sum_{A\subseteq E}\alpha_{A}\le1,\ \forall_{j\in E}\sum_{A:j\in A}\alpha_{A}\le y_{j},\ \forall_{A\subseteq E}\,\alpha_{A}\ge0}.
\]
Intuitively, the solution $\br{\alpha_{A}}{}_{A\subseteq E}$ above
represents the distribution over $2^{E}$ that maximizes the value
$\ex{f(A)}$ subject to the constraint that its marginal values satisfy
$\pr{i\in A}\le y_{i}$. The value $f^{+}\br y$ is then the expected
value of $\ex{f\br A}$ under this distribution, while the value of
$F\br y$ is the value of $\ex{f\br A}$ under the particular distribution
that places each element $i$ in $A$ independently. This relaxation
is important for our applications because the following mathematical
programming relaxation gives an upper bound on the expected value
of the optimal feasible strategy for the related stochastic probing
problem: 
\begin{equation}
\text{maximize }\setst{f^{+}\br{x\cdot p}}{x\in\P{\Iin,\Iout}}.\label{eq:mathprogram}
\end{equation}
\begin{lemma} \label{lem:math-programming-bound} Let $OPT$ be the
optimal feasible strategy for the stochastic probing problem in our
general setting, then $\ex{f\br{OPT}}\le f^{+}\br{x^{+}\cdot p}$.
\end{lemma}Proof of the following Lemma can be found in~\cite{DBLP:conf/stacs/AdamczykSW14},
and for sake of completeness we put it in the Appendix~\ref{sec:Omitted-proofs}.

However, the framework of Chekuri et al.~\cite{DBLP:conf/stoc/VondrakCZ11}
uses multilinear relaxation $F$ and not $f^{+}$. The Lemma below allows us to make a connection. 
Note that $F(x)$ is exactly equal to $\ex {F(\R)}$, i.e., it corresponds to sampling each point $e\in E$ independently with probability $x_e$, while definition of $f^+(x)$ involves the best possible, most likely correlated, distribution of $E$'s subsets. It follows immediately that for any point $x$ we have $f^+(x) \geq F(x)$, and therefore the following Lemma states
a stronger lower-bound for the measured greedy algorithm of Feldman
et al.~\cite{DBLP:conf/focs/FeldmanNS11}. 
Details are presented
in the next paragraph.

\begin{lemma} \label{lem:measured-greedy} Let $b\in\brq{0,1}$,
let $f$ be a submodular function with multilinear extension $F$,
and let $\mathcal{P}$ be any downward closed polytope. Then, the
solution $x\in\brq{0,1}^{E}$ produced by the measured greedy algorithm
satisfies 1) $x\in b\cdot{\cal P}$, 2) $F\br x\ge\br{b\cdot e^{-b}-o\br 1}\cdot\max_{y\in\mathcal{P}}f^{+}\br y$. 

\end{lemma}

\subsubsection*{Stronger bound for measured continuous greedy}

\label{sec:new-bound-measured}We now briefly review the measured
continuous greedy algorithm of Feldman et al.\ \cite{DBLP:conf/focs/FeldmanNS11}.
The algorithm runs for $\delta^{-1}$ discrete time steps, where $\delta$
is a suitably chosen, small constant. Let $y\br t$ be the algorithm's
current fractional solution at time $t$. At time $t$, the algorithm
selects vector $I\br t\in{\cal P}$ given by $\arg\max_{x\in{\cal P}}\sum_{e\in E}x_{e}\cdot\br{F\br{y\br t\lor\chr{\{e\}}}-F\br{y\br t}}$
(where $\lor$ denotes element-wise maximum). Then, it sets $y_{e}\br{t+\delta}=y_{e}\br t+\delta I_{e}(t)\cdot(1-y_{e}(t))$
and continues to time $t+\delta$.

The analysis of Feldman et al.\ shows that if, at every time step
\begin{equation}
F(y(t+\delta))-F(y(t))\ge\delta\cdot\brq{e^{-t}\cdot f\br{OPT}-F\br{y(t)}}-O\br{n^{3}\delta^{2}f\br{OPT}},\label{eq:feldman-main}
\end{equation}
then we must have $F\br{y(T)}\ge\brq{Te^{-T}-o\br 1}\cdot f(OPT).$
We note that, in fact, this portion of their analysis works even if
$f(OPT)$ is replaced by any constant value. Thus, in order to prove
our claim, it suffices to derive an analogue of \eqref{eq:feldman-main}
in which $f(OPT)$ is replaced by $f^{+}(x^{+})$, where $x^+ = \mbox{argmax}_{y\in\mathcal{P}}f^{+}\br y$. The remainder of
the proof then follows as in~\cite{DBLP:conf/focs/FeldmanNS11}.

Lemma~\ref{lem:feldman-generalization} below contains the required
analogue of \eqref{eq:feldman-main}. Hence it implies Lemma~\ref{lem:measured-greedy}.
Proof of below Lemma is placed in the Appendix~\ref{app:stronger-bound}.

\begin{lemma} \label{lem:feldman-generalization} For every time
$0\le t\le T$ 
$$F\br{y(t+\delta)}-F\br{y(t)}\ge\delta\cdot\brq{e^{-t}\cdot f^{+}(x^{+})-F(y(t))}-O(n^{3}\delta^{2})\cdot f^{+}\br{x^{+}}.$$
\end{lemma} 

\subsubsection*{Using stoch-CR schemes\label{sub:Stochastic-contention-resolution} }

The following Lemma is implied by Theorem~\ref{thm:chekurimultilinearbound}
--- it follows just from the fact that set $\act{\R}$ is distributed
exactly as $R\br{x\cdot p}$.

\begin{lemma}\label{lem:stochCRboundforF}
Let $(E,p,\Iin,\Iout)$ be a probing problem. Let $f:2^{E}\mapsto\mathbb{R}_{\geq0}$
be a non-negative submodular function with multilinear relaxation
$F$, and $x$ be a point in $b\cdot\P{\Iin,\Iout}$ for $b\in\brq{0,1}$.
Let $\bar{\pi}_{x}$ be a  $\br{b,c}$-balanced stoch-CR scheme for
$\P{\Iin,\Iout}$. Let $\bar{\pi}_{x}\br{R\br x}$ be the output of
the CR scheme, and let $\eta_{f}\br{\bar{\pi}_{x}\br{R\br x}}$ be
a pruned subset of $\bar{\pi}_{x}\br{R\br x}$. It holds that $$\ex{f\br{\eta_{f}\br{\bar{\pi}_{x}\br{R\br x}}}}\geq c\cdot F\br{x\cdot p}.$$ 

\end{lemma}

However, we cannot apply yet the above Lemma to reason about a probing
strategy, because here the pruning $\eta_{f}$ of $S$ is done after
the process. In the probing model we commit to an element once we
successfully probe it, and therefore we cannot do the pruning operation
after the execution of a stoch-CR scheme. However, since a probing
strategy inherently includes elements one-by-one, we can naturally
add to any stoch-CR scheme the pruning operation done on the fly.
The idea is to simulate the probes of elements that would be rejected
by the pruning criterion. To simulate a probe of $e$ means \textbf{not}
to probe $e$ and to toss a coin with probability $p_{e}$ --- in
case of success to behave afterwards as if $e$ was indeed taken into
the solution, and in case of failure to behave as if $e$ was not
taken. During the execution of a stoch-CR scheme $\bar{\pi}_{x}$
we construct two sets: $S^{prun}$ consists of elements successfully
probed, and $S^{virt}$ consists of elements whose simulation was
successful. If in a step we want to probe an element $e$ such that
$f\br{S^{prun}+S^{virt}+e}-f\br{S^{prun}+S^{virt}}<0$, then we simulate
the probe of $e$ and if successful $S^{virt}\leftarrow S^{virt}+e$;
otherwise we really probe $e$ and $S^{prun}\leftarrow S^{prun}+e$
if successful. We can see that at any step, it holds that $S^{prun}=\eta_{f}\br{S^{prun}+S^{virt}}$.
Also, the final random set $S^{prun}+S^{virt}$ is distributed exactly
as $\bar{\pi}_{x}\br{R\br x}$. Hence, the outcome of such a probing
strategy is $\ex{f\br{S^{prun}}}=\ex{f\br{\eta_{f}\br{S^{prun}+S^{virt}}}}=\ex{f\br{\eta_{f}\br{\bar{\pi}_{x}\br{R\br x}}}}\geq c\cdot F\br{x\cdot p},$
where the inequality comes from Lemma~\ref{lem:stochCRboundforF}.
Thus we have proven what follows.

\begin{lemma}\label{lem:pruningonthefly} Let $f$,$x$,$\bar{\pi}_{x}$
be as in Lemma~\ref{lem:stochCRboundforF}. There exists a probing
strategy whose expected \emph{outcome} is $\ex{f\br{\eta_{f}\br{\bar{\pi}_{x}\br{R\br x}}}}\geq c\cdot F\br{x\cdot p}$.

\end{lemma}Now we can finish the proof of Theorem~\ref{thm:maintheorem}.
Consider the following algorithm. First, use Lemma~\ref{lem:measured-greedy}
to find a point $x^{*}$ such that
$$F\br{x^{*}\cdot p}\geq\br{b\cdot e^{-b}-o\br 1}\cdot\max_{x\in\P{\Iin,\Iout}}f^{+}\br{x\cdot p}.$$
Second, run the probing strategy based on a stoch-CR scheme $\bar{\pi}_{x^{*}}$
as described in Lemma~\ref{lem:pruningonthefly}. This yields, together
with Lemma~\ref{lem:math-programming-bound}, that the outcome of
such a probing strategy is
\begin{multline*}
\ex{f\br{\eta_{f}\br{\bar{\pi}_{x^{*}}\br{R\br{x^{*}}}}}}\geq c\cdot F\br{x^{*}\cdot p}\geq c\cdot\br{b\cdot e^{-b}-o\br 1}\cdot\max_{x\in\P{\Iin,\Iout}}f^{+}\br{x\cdot p}\\
\geq c\cdot\br{b\cdot e^{-b}-o\br 1}\cdot\ex{f\br{OPT}}.
\end{multline*}

In~\cite{DBLP:conf/stoc/VondrakCZ11} also an alternative approach than pruning was used.
They defined a \emph{strict} contention resolution scheme where the approximation guarantee $\pr{e\in \pi_x\br{\R} | e\in \R} \geq c$ holds with equality rather than inequality. Since the pruning operation
depends on an objective function, resigning from it allows for the algorithm to be used in maximizing many submodular functions at the same time.
In our stochastic setting we can also skip the pruning operation if we have a strict scheme.
No proof of this fact is needed, since we can directly use an appropriate analog of Lemma~\ref{lem:stochCRboundforF} (Theorem~4.1 from~\ref{thm:chekurimultilinearbound}).

\subsubsection*{Stochastic probing for various constraints\label{sub:Stochastic-probing-for-varous-constraints}}

Gupta and Nagarajan~\cite{DBLP:conf/ipco/GuptaN13} introduced a
notion of \emph{ordered CR scheme}, for which there exists a (possibly
random) permutation $\sigma$ of $E$, so that for each $A$ the set
$\pi_{x}\br A$ is the maximal independent subset of $\I$ obtained
by considering elements in the order of $\sigma$.
Ordered scheme are required to implement probing strategies, because of the commitment to the elements.
CR schemes exist for various types of constraints, e.g., matroids, sparse packing integer programs, constant number of knapsacks, unsplittable flow on trees (UFP), $k$-systems (including intersection of $k$ matroids, and $k$-matchoids). Ordered schemes exist for k-systems and UFP on trees. See Theorem~4 in~\cite{DBLP:conf/ipco/GuptaN13} for a listing with exact parameters.

The following Lemma is based on Lemma~1.6 from~\cite{DBLP:conf/stoc/VondrakCZ11}.
The proof basically carries over, the only thing we have to do is to again incorporate
probes' simulations as in the proof of Lemma~\ref{lem:pruningonthefly}. Proof of Lemma~\ref{lem:combining} is placed in Appendix~\ref{sec:Omitted-proofs}.
Theorem~3.4 from~\cite{DBLP:conf/ipco/GuptaN13} yields a similar result but would imply a 
 $\br{b,c_{out} + c_{in} - 1}$-balanced stoch-CR scheme. Thus the below Lemma can be considered 
 as a strengthening of Theorem~3.4 from~\cite{DBLP:conf/ipco/GuptaN13}, because an FKG inequality is used in the proof instead of a union-bound.

\begin{lemma}
\label{lem:combining}Consider a probing problem $\br{E,p,\Iin,\Iout}$.
Suppose we have a $\br{b,c_{out}}$-balanced CR-scheme $\pi^{out}$
for $\P{\Iout}$, and a $\br{b,c_{in}}$-balanced\textbf{ }ordered
CR scheme $\pi^{in}$ for $\P{\Iin}$. Then there exists a $\br{b,c_{out}\cdot c_{in}}$-balanced
stoch-CR scheme for $\P{\Iin,\Iout}$.
\end{lemma}

In light of the above Lemma, one can question Definition~\ref{def:stochCRscheme} of a stoch CR-scheme ---
why do we need to define it at all, if we can just be using two separate classic CR schemes $\pi^{out}$, $\pi^{in}$. The reason is that there may exist stoch-CR schemes that are not convolutions of two deterministic schemes, and that yield better approximations than corresponding convoluted ones. Such a stoch-CR scheme is presented in Section~\ref{sec:Stochastic-contention-resolution}.

In a recent paper, Feldman et al.~\cite{DBLP:journals/corr/FeldmanSZ15} presented a variant of online contention resolution schemes. They enriched the set of constraints possible to use in the stochastic probing problem --- previously inner knapsack constraints were not possible to incorporate, as well as deadlines (element $e$ can be taken only first $d_e$ steps) for weighted settings.
Their results can be extended to monotone submodular settings by making use of 
a stronger bound for continuous greedy algorithm~\cite{Calinescu2011} presented
in~\cite{DBLP:conf/stacs/AdamczykSW14}. The stronger bound for measured greedy algorithm --- which
works for non-monotone functions --- that we give
in this paper can also be used in~\cite{DBLP:journals/corr/FeldmanSZ15} enhancing their result
by the possibility of handling non-monotone functions as well.   

\section{Stoch-CR scheme for intersection of transversal matroids\label{sec:Stochastic-contention-resolution}}

In this section we prove Lemma~\ref{lem:stochCRscheme}. We assume
we have only one inner matroid $\M^{in}$ to keep the presentation
simple. Also, we shall present a $\br{1,\frac{1}{2}}$-balanced CR
scheme, instead of $\br{b,\frac{1}{b+1}}$. In the Appendix~\ref{sec:Full-proof-of}
we present a full scheme with arbitrary $b,\kin,\mbox{ and }\kout$.

Our stoch-CR scheme on the input is given a point $x$ such that $p\cdot x\in\P{\M^{in}}$
and a set $A\subseteq E$, and on the output it returns set $\bar{\pi}_{x}\br A$.
The procedure is divided in two phases. First, the preprocessing phase,
depends only on the point $x$. Second, the random selection phase,
depends on the set $A\subseteq E$ and the outcome of the preprocessing
phase.

\subsubsection*{Matroid properties}

Let $\M^{in}=\br{E,\I^{in}}$. We know~\cite{Schrijver:book} that
the convex hull of $\setst{\chr A}{A\in{\cal I}}$, i.e.,\ characteristic
vectors of all independent sets of $\M^{in}$, is equivalent to the
\emph{matroid polytope} ${\cal P}\br{\M^{in}}=\setst{x\in\mathbb{R}_{\geq0}^{E}}{\forall_{A\in{\cal I}^{in}}\sum_{e\in A}p_{e}\cdot x_{e}\leq r_{\M^{in}}\br A}$,
where $r_{\M^{in}}$ is the rank function of $\M^{in}$. Thus for
any $x\in{\cal P}\br{\M^{in}}$ in polynomial time we can find representation
$p\cdot x=\sum_{i=1}^{m}\beta_{i}\cdot\chr{B_{i}}$, where $B_{1},\ldots,B_{m}\in{\cal I}$
and $\beta_{1},\ldots,\beta_{m}$ are non-negative weights such that
$\sum_{i=1}^{m}\beta_{i}=1$ . We shall call sets $B_{1},\ldots,B_{m}$
the \emph{support }of $p\cdot x$ in $\M^{in}$, and denote it by
${\cal B}$.

In the remainder of the section we assume that we know the graph representation
$\br{E\cup V,\subseteq E\times V}$ of the matroid described below.
This assumption is quite natural and common, e.g.,~\cite{DBLP:conf/soda/BabaioffIK07}.
\emph{}

\begin{definition} Consider bipartite graph $\br{E\cup V,\subseteq E\times V}$.
Let $\I$ be a family of all subsets $S$ of $E$ such that there
exists a matching between $S$ and $V$ of size exactly $|S|$. Then $M=\br{E,\I}$ is a matroid,
called \emph{transversal }matroid. \end{definition}

\subsubsection*{Preprocessing}

In what follows we shall write superscripts indicating the time in
which we are in the process.

We start by finding the support $\B^{0}$ of vector $p\cdot x\in\P{\M^{in}}$,
i.e., $p\cdot x=\sum_{i}\beta_{i}\cdot\chr{B_{i}^{0}}$. For every
two sets $B,A\in{\cal B}^{0}$ we find a mapping $\phi^{0}[B,A]:B\rightarrow A\cup\{\perp\}$,
which we call \emph{transversal mapping.} This mapping will satisfy
three properties.

\begin{property}

For each $a\in A$ there is at most one $b\in B$ for which $\phi^{0}[B,A]\br b=a$.

\end{property}

\begin{property}

For $b\in B\setminus A$, if $\phi^{0}[B,A]\br b=\bot$, then $A+b\in{\cal I}$,
otherwise $A-\phi^{0}[B,A]\br b+b\in{\cal I}$. 

\end{property}Note that unlike in standard exchange properties of
matroids, we do not require that $\phi^{0}\brq{B,A}\br b=b$, if $b\in A\cap B$.
Property 3 will be presented in a moment. Once we find the family
$\phi^{0}$ of transversal mappings, for each element $e\in E$ we
choose one set among $B_{i}^{0}:e\in B_{i}^{0}$ with probability
$\beta_{i}/p_{e}x_{e}$; since $\sum_{B_{i}^{0}:e\in B_{i}^{0}}\beta_{i}=p_{e}x_{e}$
this is properly defined. Denote by $c\br e$ the index of the chosen
set, and call \emph{$e$-critical} the set \emph{$B_{c\br e}^{t}$}
for any $t$ (note that $c(e)$ is fixed throughout the
process). We concisely denote indices of critical sets by ${\cal C}=\br{c\br e}_{e\in E}$.
For each element $e$ we define $\Gamma^{0}\br e=\setst f{f\neq e\wedge\phi^{0}\brq{B_{c\br f}^{0},B_{c\br e}^{0}}\br f=e}$
--- the \emph{blocking set }of\emph{ }$e$. The Lemma below follows
from Property 1, and its proof is in the Appendix~\ref{sec:Omitted-proofs}.

\begin{lemma}\label{lem:blockingset}If $p\cdot x\in{\cal P}\br{\M^{in}}$,
then for any element $e$, it holds that\newline $\exls{{\cal C},R\br x}{\sum_{f\in\Gamma^{0}\br e}p_{f}\cdot\indi{f\in R\br x}}\leq1$,
where the expectation is over $R\br x$ and the choice of critical
sets ${\cal C}$; here $\indi{{\cal E}}$ is a 0-1 indicator of random
event ${\cal E}$.

\end{lemma}

\subsubsection*{Random selection procedure}

\begin{algorithm}
\caption{\label{alg:Random-selection-phase}Stoch-CR scheme $\bar{\pi}_{x}\protect\br A$}

\begin{algorithmic}[1]

\STATE find support $\B^{0}$ of $p\cdot x$ in $\M^{in}$ and family
$\phi^{0}$; choose critical sets ${\cal C}$

\STATE remove from $A$ all $e:$ $x_{e}=0$; mark all $e\in A$
as available

\STATE $S\leftarrow\emptyset$

\STATE \textbf{while} there are still available elements in $A$
\textbf{do\label{alglabel:while}}

\STATE $\qquad$pick element $e$ uniformly at random from $A$\label{alglabel:pick-element}

\STATE $\qquad$\textbf{if }$e$ is available \textbf{then}

\STATE $\qquad$$\qquad$probe $e$

\STATE $\qquad$$\qquad$\textbf{if }probe of $e$ successful \textbf{then }

\STATE $\qquad$$\qquad$$\qquad$$S\leftarrow S\cup\set e$

\STATE $\qquad$$\qquad$$\qquad$\textbf{for }each set $B_{i}^{t}$
of support $\B^{t}$ \textbf{do}

\STATE $\qquad$$\qquad$$\qquad$$\qquad$$B_{i}^{t}\leftarrow B_{i}^{t}+e$

\STATE $\qquad$$\qquad$call $e$ unavailable

\STATE $\qquad$\textbf{else} simulate the probe of $e$

\STATE $\qquad$\textbf{if }probe or simulation was successful \textbf{then}

\STATE $\qquad$$\qquad$\textbf{for }each set $B_{i}^{t}$ of support
$\B^{t}$ \textbf{do\label{alglabel:supportupdate}}

\STATE $\qquad$$\qquad$$\qquad$$f\leftarrow\setphi{B_{c\br e}^{t}}{B_{c\br f}^{t}}e$

\STATE $\qquad$$\qquad$$\qquad$\textbf{if} $f\neq e$ \textbf{then
}$B_{c\br f}^{t}\leftarrow B_{c\br f}^{t}-f$ and call $f$ unavailable\label{alglabel:remove-element}

\STATE $\qquad$compute the family $\phi^{t+1}$ 

\STATE $\qquad$\textbf{for} each $i$ \textbf{do} $B_{i}^{t+1}\leftarrow B_{i}^{t}$ 

\STATE $\qquad$$t\leftarrow t+1$; 

\STATE \textbf{return $S$ }as \textbf{$\bar{\pi}_{x}\br A$}

\end{algorithmic}
\end{algorithm}

The whole stoch-CR scheme is presented in Algorithm~\ref{alg:Random-selection-phase}.
During the algorithm we modify sets of support $\B^{t}$ after each
step, but we keep the weights $\beta_{i}$ unchanged. We preserve
an invariant that each $B_{i}^{t}\mbox{ from }\B^{t}$ is an independent
set of matroid $\M^{in}$. At the end of the algorithm the set $\bar{\pi}_{x}\br A$
belongs to every set $B_{i}^{t}\in\B^{t}$. Hence, the final set $\bar{\pi}_{x}\br A$
is independent in every matroid. 

Now we define Property 3 of transversal mappings. Suppose that in
the first step we update the support $\B^{0}$ according to the for
loop in line~\ref{alglabel:supportupdate}, and we obtain $\B^{1}$.
Different support $\B^{1}$ requires a different family of mappings
$\phi^{1}$, and so in step 2, the elements that can block $e$ are
$\Gamma^{1}\br e$. If it happens that $\Gamma^{1}\br e\neq\Gamma^{0}\br e$,
then we cannot show the monotonicity property of stoch-CR scheme.
However, we can require from the transversal mappings to keep the
blocking sets $\Gamma^{t}\br e$ unchanged, as long as $e$ is available.
In the Appendix~\ref{sec:Transversal-mappings} we show how to find
such a family of transversal mappings $\phi^{0}$ and how to construct
$\phi^{t+1}$ given $\phi^{t}$.

\begin{property}

Let $\phi^{t}$ be a family of transversal mappings for $\B^{t}$.
Suppose we update the support $\B^{t}$ and obtain $\B^{t+1}$. Then
we can find a family $\phi^{t+1}$ of transversal mappings such that
$\Gamma^{t}\br e=\Gamma^{t+1}\br e$ for any element $e$ that is
still available after step $t$.

\end{property}

\subsubsection*{Analysis}

First, an explanation. We allow to pick in line~\ref{alglabel:pick-element}
elements that we have once probed and simulate their probe. This guarantees
that the probability that an available element is blocked is equal
for every step. Otherwise, again, we would not be able to guarantee
the monotonicity.

In the analysis we deploy martingale theory. In particular Doob's
Stopping Theorem which states that if a martingale $\left(Z^{t}\right)_{t\geq0}$
and stopping time $\tau$ are both ``well-behaving'', then $\ex{Z^{\tau}}=\ex{Z^{0}}$.
In the Appendix~\ref{sec:Martingale-Theory} we put necessary definitions,
statement of Doob's Stopping Theorem, and an explanation of why we
can use this Theorem. 

The random process executed in the while loop depends on the critical
sets chosen in the preprocessing phase. Therefore, when we analyze
the random process we condition on the choice of critical sets ${\cal C}$.

We say $e$ is \emph{available,} if it is still possible to probe,
i.e., it is not yet blocked, and it was not yet probed. Define $X_{e}=\brqbb{e\in A}$;
in Iverson notation $\brqbb{false}=0=1-\brqbb{true}$.

Let $Y_{e}^{t}$ for $t=0,1,...$, be a random variable indicating
if $e$ is still available after step $t$. Initially $Y_{e}^{0}=X_{e}$.
Let $P_{e}^{t}$ be a random variable indicating if $e$ was probed
in one of steps $0,1,...,\mbox{ or }t$; we have $P_{e}^{0}=0$ for
all $e$. Variable $P_{e}^{t+1}-P_{e}^{t}$ indicates if $e$ was
probed at step $t+1$. Given the information ${\cal F}^{t}$ about
the process up to step $t$, the probability of this event is $$\excond{P_{e}^{t+1}-P_{e}^{t}}{{\cal F}^{t},{\cal C}}=\frac{Y_{e}^{t}}{\size A},$$
because if element $e$ is still available after step $t$ (i.e.,
$Y_{e}^{t}=1$), then with probability $\frac{1}{\size A}$ we choose
it in line~\ref{alglabel:pick-element}, and otherwise (i.e. $Y_{e}^{t}=0$)
we cannot probe it.

Variable $Y_{e}^{t}-Y_{e}^{t+1}$ indicates whether element $e$ stopped
being available at step $t+1$, i.e., we either have picked it in
line~\ref{alglabel:pick-element} and probed (with probability $\frac{Y_{e}^{t}}{\size A}$),
or some $f\in\Gamma^{t}\br e$ has blocked $e$ in line~\ref{alglabel:remove-element}
(with probability $\frac{Y_{e}^{t}}{\size A}\cdot\sum_{f\in\Gamma^{t}\br e}p_{f}X_{f}$).
So in total $$\excond{Y_{e}^{t}-Y_{e}^{t+1}}{{\cal F}^{t},{\cal C}}=\frac{Y_{e}^{t}}{\size A}\br{1+\sum_{f\in\Gamma^{t}\br e}p_{f}X_{f}},$$
and we can say that
$$\excond{\br{P_{e}^{t+1}-P_{e}^{t}}\cdot\br{1+\sum_{f\in\Gamma^{t}\br e}p_{f}X_{f}}-\br{Y_{e}^{t}-Y_{e}^{t+1}}}{{\cal F}^{t},{\cal C}}=0.$$

This means that the sequence $\br{\br{1+\sum_{f\in\Gamma^{t}\br e}p_{f}X_{f}}\cdot P_{e}^{t}+Y_{e}^{t}}_{t\geq0}$
is a martingale. Let $\tau=\min\setst t{Y_{e}^{t}=0}$ be the step
in which edge $e$ became unavailable. It is clear that $\tau$ is
a stopping time (definition in Appendix~\ref{sec:Martingale-Theory}).
Thus from Doob's Stopping Theorem we get that
$$\excondls{\tau}{\br{1+\sum_{f\in\Gamma^{\tau}\br e}p_{f}X_{f}}\cdot P_{e}^{\tau}+Y_{e}^{\tau}}{{\cal C}}=\excondls{}{\br{1+\sum_{f\in\Gamma^{0}\br e}p_{f}X_{f}}\cdot P_{e}^{0}+Y_{e}^{0}}{{\cal C}},$$
and this is equal to $X_{e}$, because $P_{e}^{0}=0$ and $Y_{e}^{0}=X_{e}$.
We have $Y_{e}^{\tau}=0$, and expression $1+\sum_{f\in\Gamma^{\tau}\br e}p_{f}X_{f}$
is in fact equal to $1+\sum_{f\in\Gamma^{0}\br e}p_{f}X_{f}$ (Property
3 of the transversal mapping, as $e$ was available before step $\tau$),
which depends solely on ${\cal C}$ and $A$.
Hence $$\br{1+\sum_{f\in\Gamma^{0}\br e}p_{f}X_{f}}\cdot\excondls{\tau}{P_{e}^{\tau}}{{\cal C}}=X_{e}.$$
Now just note that $\excondls{\tau}{P_{e}^{\tau}}{{\cal C}}$ is exactly
the probability that $e$ is probed, so we conclude that $$\prcond{e\in\bar{\pi}_{x}\br A}{{\cal C}}=p_{e}\cdot\excondls{\tau}{P_{e}^{\tau}}{{\cal C}}=p_{e}X_{e}\left/\br{1+\sum_{f\in\Gamma^{0}\br A}p_{f}X_{f}}.\right.$$

\paragraph{Monotonicity}

Set $\Gamma^{0}\br e$ does not depend on $A$, but only on the vector
$p\cdot x$ and ${\cal C}$, so for $A_{1}\subseteq A_{2}$ we have
$\sum_{f\in\Gamma^{0}\br A}p_{f}\cdot\brqbb{f\in A_{1}}\leq\sum_{f\in\Gamma^{0}\br A}p_{f}\cdot\brqbb{f\in A_{2}}.$

\paragraph{Approximation guarantee}

In the identity $$\prcond{e\in\bar{\pi}_{x}\br A}{{\cal C}}=p_{e}X_{e}\left/\br{1+\sum_{f\in\Gamma^{0}\br A}p_{f}X_{f}}\right.$$
we place random set $\R$ instead of $A$; now $X_{f}=\indi{f\in R\br x}$
is a random variable. Let us condition on $e\in R\br x$, take expected
value $\excondls{{\cal C},R\br x}{\cdot}{e\in R\br x}$, and apply
Jensen's inequality to convex function $x\mapsto\frac{1}{x}$ to get:
\begin{align*}
\prcond{e\in\bar{\pi}_{x}\br{\R}}{e\in R\br x}&=\excondls{{\cal C},R\br x}{\prcond{e\in\bar{\pi}_{x}\br{\R}}{{\cal C},R\br x}}{e\in R\br x}\\
&=\excondls{{\cal C},R\br x}{p_{e}X_{e}\left/\br{1+\sum_{f\in\Gamma^{0}\br e}p_{f}X_{f}}\right.}{e\in R\br x}
\hfill\\
&\geq p_{e}\left/\excondls{{\cal C},R\br x}{1+\sum_{f\in\Gamma^{0}\br e}p_{f}X_{f}}{e\in R\br x}.\right.
\end{align*}
Since $\excondls{{\cal C},R\br x}{\sum_{f\in\Gamma^{0}\br e}p_{f}X_{f}}{e\in R\br x}\leq1$
from Lemma~\ref{lem:blockingset}, we conclude that $$\prcond{e\in\bar{\pi}_{x}\br{\R}}{e\in R\br x}\geq\frac{p_{e}}{2},$$
and therefore $\prcond{e\in\bar{\pi}_{x}\br{\R}}{e\in\act{R\br x}}\geq\frac{1}{2}$,
which is exactly Property 3 from the definition of stoch-CR scheme.

\subsection*{Acknowledgments}

I thank Justin Ward for his help in proving Lemma~\ref{lem:feldman-generalization},
and also for valuable suggestions that helped to improve the presentation
of the paper.

\newpage{}
\bibliographystyle{plain}
\bibliography{stochasticresolution}

\newpage{}

\appendix

\section{\label{app:stronger-bound}Stronger bound for measured continuous
greedy}

Recall, we need to prove Lemma~\ref{lem:feldman-generalization}.
To do so, we shall require the following additional facts from the
analysis of \cite{DBLP:conf/focs/FeldmanNS11}.\\
\begin{lemma}[Lemma 3.3 in \cite{DBLP:conf/focs/FeldmanNS11}]
\label{lem:feldman-1-1} Consider two vectors $x,x'\in[0,1]^{E}$,
such that for every $e\in E$, $\size{x_{e}-x'_{e}}\le\delta$. Then,
$F(x')-F(x)\ge\sum_{e\in E}(x'_{e}-x_{e})\cdot\partial_{e}F(x)-O\br{n^{3}\delta^{2}}\cdot f\br{OPT}$.

\end{lemma}

\begin{lemma}[Lemma 3.5 in \cite{DBLP:conf/focs/FeldmanNS11}] \label{lem:feldman-2-1}
Consider a vector $x\in\brq{0,1}{}^{E}$. Assuming $x_{e}\le a$ for
every $e\in E$, then for every set $S\subseteq E$, $F\br{x\lor\chr S}\ge(1-a)f\br S$. 

\end{lemma}

\begin{lemma}[Lemma 3.6 in \cite{DBLP:conf/focs/FeldmanNS11}] \label{lem:feldman-3-1}
For every time $0\le t\le T$ and element $e\in E$, $y_{e}\br t\le1-\br{1-\delta}{}^{t/\delta}\le1-\exp\br{-t}+O\br{\delta}$.

\end{lemma}\textbf{}\\
\textbf{Lemma~\ref{lem:feldman-generalization}.} For every time
$0\le t\le T$:

$F\br{y(t+\delta)}-F\br{y(t)}\ge\delta\cdot\brq{e^{-t}\cdot f^{+}(x^{+})-F(y(t))}-O(n^{3}\delta^{2})\cdot f^{+}\br{x^{+}}$. 

\begin{proof} Applying Lemma \ref{lem:feldman-1-1} to the solutions
$y(t+\delta)$ and $y(t)$, we have 
\begin{align}
 & F\br{y(t+\delta)}-F(y(t))\nonumber \\
\ge & \sum_{e\in E}\delta\cdot I_{e}\br t(1-y(t))\cdot\partial_{j}F(y(t))-O\br{n^{3}\delta^{2}})\cdot f(OPT)\\
= & \sum_{e\in E}\delta\cdot I_{e}\br t(1-y(t))\cdot\frac{F\br{y(t)\lor\chr j}-F(y(t))}{1-y(t)}-O(n^{3}\delta^{2})\cdot f(OPT)\nonumber \\
= & \sum_{e\in E}\delta\cdot I_{e}\br t\cdot\brq{F(y(t)\lor\chr j)-F(y(t))}-O(n^{3}\delta^{2})\cdot f(OPT)\nonumber \\
\ge & \sum_{e\in E}\delta\cdot x_{e}^{+}\brq{F(y(t)\lor\chr j)-F(y(t))}-O(n^{3}\delta^{2})\cdot f(OPT)\label{eq:cg-1-1}
\end{align}
where the last inequality follows from our choice of $I(t)$.

Moreover, we have $f^{+}(x^{+})=\sum_{A\subseteq E}\alpha_{A}f(A)$
for some set of values $\alpha_{A}$ satisfying $\sum_{A\subseteq E}\alpha_{A}=1$
and $\sum_{A\subseteq E:e\in A}\alpha_{A}=x_{e}^{+}$. Thus, 
\begin{multline*}
\sum_{e\in E}x_{e}^{+}\brq{F(y(t)\lor\chr j)-F(y(t))}=\sum_{A\subseteq E}\alpha_{A}\sum_{j\in A}\brq{F(y(t)\lor\chr j)-F(y(t))}\\
\ge\sum_{A\subseteq E}\alpha_{A}\brq{F(y(t)\lor\chr A)-F(y(t))}\ge\sum_{A\subseteq E}\alpha_{A}\brq{(e^{-t}-O(\delta))\cdot f(A)-F(y(t))}\\
=(e^{-t}-O(\delta))\cdot f^{+}(x^{+})-F(y(t)).
\end{multline*}
where the first inequality follows from the fact that $F$ is concave
in all positive directions, and the second from Lemmas \ref{lem:feldman-2-1}
and \ref{lem:feldman-3-1}. Combining this with the inequality \eqref{eq:cg-1-1},
and noting that $f^{+}\br{x^{+}}\ge f^{+}\br{OPT}=f\br{OPT}$, we
finally obtain $F(y(t+\delta))-F(y(t))\ge\delta\cdot\brq{e^{-t}\cdot f^{+}(x^{+})-F(y(t))}-O(n^{3}\delta^{2})\cdot f^{+}(x^{+}).$
\end{proof}

\section{\label{sec:Transversal-mappings}Transversal mappings}

Recall the definition of a transversal matroid.\begin{definition}
Consider a bipartite graph $\br{E\cup V,\subseteq E\times V}$. Let
$\I$ be a family of all subsets $S$ of $E$ such that there exists
an injection from $S$ to $V$. Then $M=\br{E,\I}$ is a matroid,
called \emph{transversal }matroid. \end{definition}

We assume we know the graph $\br{E\cup V,\subseteq E\times V}$ of
the matroid.

Let $\B^{0}$ be the initial support. Let $A\in\B^{0}$ be an independent
set. From the definition of the transversal matroid, there exists
an injection $v^{A}:A\rightarrow V$; we shall say that $a\in A$
is \emph{matched to }$v^{A}\br a$. There can be many such injections
for a given set, but we initially pick one for every $A\in\B^{0}$.
When a set of the support will be changed we shall explicitly define
an injection. In fact, only for added elements we will define a new
match, for all other elements they will be matched all the time to
the same vertex, as long as they are available.

For any two $A,B\in\B^{0}$ we define the mapping $\phi^{0}[B,A]:B\rightarrow A\cup\{\perp\}$
as follows. Let $v^{A}$ be the injection of $A$, and let $v^{B}$
be the injection of $B$. If there exists element $a$ such that $v^{A}\br a=v^{B}\br b$,
then we set $\phi^{0}[B,A]\br b=a$; if not, we set $\phi^{0}[B,A]\br b=\perp$.

Let us verify that such a definition satisfies first two properties.\\
\textbf{Property 1.}\emph{ For each $a\in A$ there is at most one
$b\in B$ for which $\phi^{0}[B,A]\br b=a$.}

This one is trivially satisfied because there can be at most one $b\in B$
that according to $v^{B}$ is matched to $v^{A}\br a$.$\hfill\square$\\
\\
\textbf{Property 2.} \emph{For $b\in B\setminus A$, if $\phi^{0}[B,A]\br b=\bot$,
then $A+b\in{\cal I}$, otherwise $A-\phi^{0}[B,A]\br b+b\in{\cal I}$.}

Suppose $\phi^{0}[B,A]\br b=\bot$. It means that $b$ is matched
to $v^{B}\br b$ to which no element $a\in A$ is matched to. Therefore
when we add edge $(b,v^{B}\br b)$ to the injection $\set{\br{a,v^{A}\br a}}_{a\in A}$
it is still a proper injection, since $b\notin A$, and so $A+b\in\I$.
Now suppose $\phi^{0}[B,A]\br b=a'\neq\perp$. Now the set $A$ changes
to $A-a'+b$ and the underlying injection is $\set{\br{a,v^{A}\br a}}_{a\in A\setminus a'}\cup\{(b,v^{B}\br b)\}$,
which is a valid injection since $b\notin A$, and if so, then $A-a'+b$
is indeed an independent set. So Property 2 also holds.$\hfill\square$\\

Now let us move to the most technically demanding property.\emph{}\\
\textbf{Property 3.}\emph{ Let $\phi^{t}$ be a family of transversal
mappings for $\B^{t}$. Suppose we update the support $\B^{t}$ and
obtain $\B^{t+1}$. Then we can find a family $\phi^{t+1}$ of transversal
mappings such that $\Gamma^{t}\br a=\Gamma^{t+1}\br a$ for any element
$a$ that is still available after step $t$.}

Suppose that in step $t$ we have chosen element $c$, and we update
the support $\B^{t}$ as described in the for loop of the algorithm
in line~\ref{alglabel:supportupdate}. First of all assume that $c\neq a$,
otherwise $a$ becomes unavailable so there is nothing to prove. Let
$C^{t}$ be the critical set of $c$ and let $v^{C^{t}}\br c$ be
the vertex to which $c$ is matched according to $v^{C^{t}}$. Consider
set $B^{t}\in\B^{t}$ and let us describe how $c$ affects $B^{t+1}$
and injection $v^{B^{t+1}}$. 

\begin{figure}[H] 
\begin{center}
\subfloat{
\scalebox{0.33}{
\input{injectionUpdate1.pstex_t}
}
}
\caption{\label{fig:injectionUpdate1}Illustration of Case 1, $c\in B^t$}
\end{center}
\end{figure}
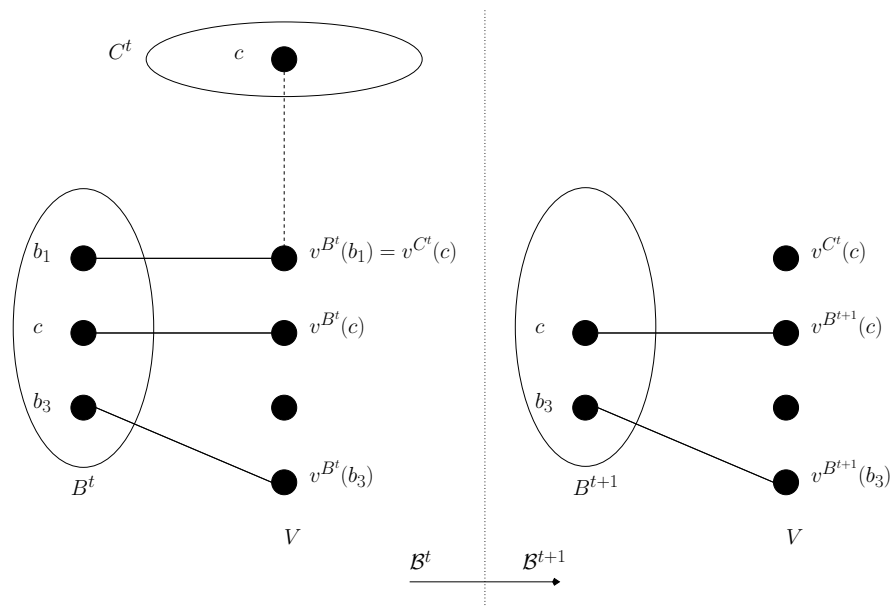

\emph{Case 1, $c\in B^{t}:$ }If it is $c$ from $B^{t}$ that is
matched to $v^{C^{t}}\br c$, i.e., $v^{B^{t}}\br c=v^{C^{t}}\br c$,
then we do not have to change anything, we set $B^{t+1}:=B^{t}$ and
$v^{B^{t+1}}=v^{B^{t}}$. If $v^{B^{t}}\br c\neq v^{C^{t}}\br c$,
then let $b_{1}$ be such that $v^{B^{t}}\br{b_{1}}=v^{C^{t}}\br c$.
We remove $b_{1}$ from $B^{t}$, i.e., $B^{t+1}:=B^{t}\setminus b_{1}$
(we do not have to add $c$ to $B^{t+1}$ because it is already there).
For every $b_{3}\in B^{t+1}$ we set $v^{B^{t+1}}\br{b_{3}}=v^{B^{t}}\br{b_{3}}$.
See Figure~\ref{fig:injectionUpdate1}.

\newpage

\begin{figure}[H]
\begin{center}
\subfloat{
\scalebox{0.33}{
\input{injectionUpdate2.pstex_t}
}
}
\caption{\label{fig:injectionUpdate2}Illustration of Case 2, $c\notin B^t$}
\end{center}
\end{figure}
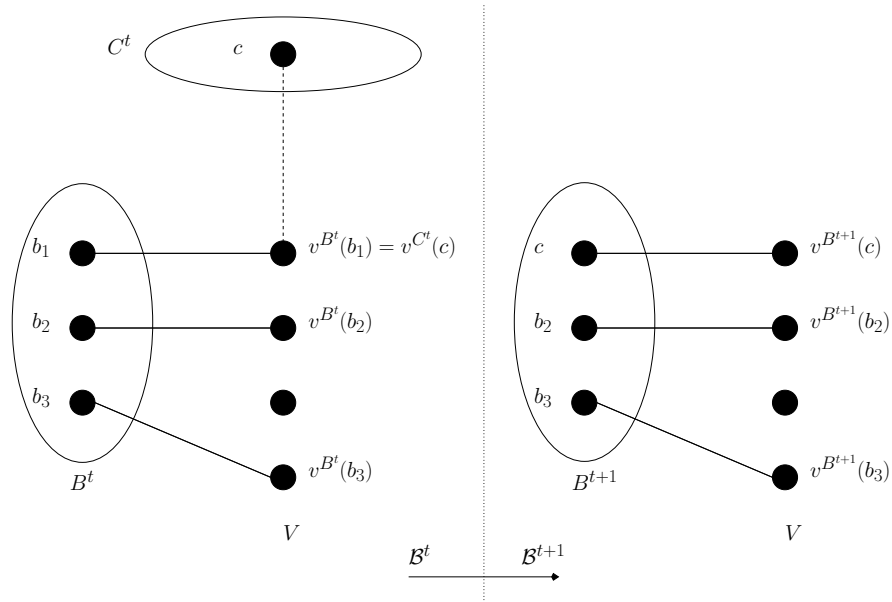

\emph{Case 2, $c\notin B^{t}:$} Let $b_{1}$ be such that $v^{B^{t}}\br{b_{1}}=v^{C^{t}}\br c$.
We remove $b_{1}$ from $B^{t}$ and add $c$ instead, i.e., $B^{t+1}=B^{t}-b_{1}+c$.
The injection is defined as: $v^{B^{t+1}}\br c=v^{C^{t}}\br c$, and
$v^{B^{t+1}}\br{b_{3}}=v^{B^{t}}\br{b_{3}}$ for $b_{3}\in B^{t}\setminus b_{1}$.
See Figure~\ref{fig:injectionUpdate2}.

\newpage

Given sets $A^{t+1},B^{t+1}$ with corresponding injections $v^{A^{t+1}}$,$v^{B^{t+1}}$
we define mapping $\phi^{t+1}$ as before, i.e., $\phi^{t+1}\brq{B^{t+1},A^{t+1}}\br b=a$,
if $v^{A^{t+1}}\br a=v^{B^{t+1}}\br b$. 

Now we need to show that the set $\Gamma^{t+1}\br a=\setst b{b\neq a\wedge\setphi{B_{c\br b}^{t+1}}{B_{c\br a}^{t+1}}b=a}$
is equal to $\Gamma^{t}\br a=\setst b{b\neq a\wedge\setphi{B_{c\br b}^{t}}{B_{c\br a}^{t}}b=a}$,
if $a$ is still available.

Consider again sets $A^{t+1},B^{t+1}$ and suppose that $A^{t}$ is
the critical set of $a$ and that $B^{t}$ is the critical set of
$b$. Suppose that both $a,b$ are matched to $v_{ab}=v^{A^{t}}\br a=v^{B^{t}}\br b$,
i.e., $b\in\Gamma^{t}\br a$. If it happened that in step $t$ element
$c$ removed $a$ and $b$, i.e., $v^{C^{t}}\br c=v_{ab}$, then elements
$a,b$ are blocked and not available, so there is nothing to prove
here. If $v^{C^{t}}\br c\neq v$, then from the reasoning in Case
1 and 2, we know that $a$ and $b$ are still matched to $v_{ab}$,
i.e., $v_{ab}=v^{A^{t+1}}\br a=v^{B^{t+1}}\br b$. But if so, then
$\phi^{t+1}\brq{B^{t+1},A^{t+1}}\br b=a$, and $b\in\Gamma^{t+1}\br a$
still, because $A^{t+1},B^{t+1}$ remain critical sets of $a,b$.
Conversely, if $b$ is not matched to $v^{A^{t}}\br a$, i.e., $v^{B^{t}}\br b\neq v^{A^{t}}\br a$,
then $b\notin\Gamma^{t}\br a$. But if $c$ during the update does
not block $b$, then $b$ does not change its matched vertex so we
still have $v^{B^{t+1}}\br b\neq v^{A^{t+1}}\br a$, and still $b\notin\Gamma^{t+1}\br a$.

\begin{figure}[H]
\begin{center}
\subfloat{
\scalebox{0.25}{
\input{gammaUpdate.pstex_t}
}
}
\caption{\label{fig:gammaUpdate}Illustration of change in $\Gamma$. We have blocked $b_1$, and if $b_1 \in \Gamma^t(a_1)$, then it does not matter anyway, because we have also blocked $a_1$. Element $c$ was matched (w.r.t. $B_t$) to the same vertex as $a_2$, but $B_t$ is not the critical set of $c$, so $c\notin \Gamma^t(a_2)$. Assume $B^t$ is a critical set of $b_3$: we have $\phi^t[B^t,A^t](b_3)=a_3$ and so $b_3\in \Gamma^t(a_3)$; after the update we still have $\phi^{t+1}[B^{t+1},A^{t+1}](b_3)=a_3$, so $b_3 \in \Gamma^{t+1}(a_3)$. Element $a_4$ did not have any element $b'\in B^t$ in $\Gamma^t(a_4)$, so it does not have any $b' \in B^{t+1}$ in $\Gamma^{t+1}(a_4)$ as well.}
\end{center}
\end{figure}
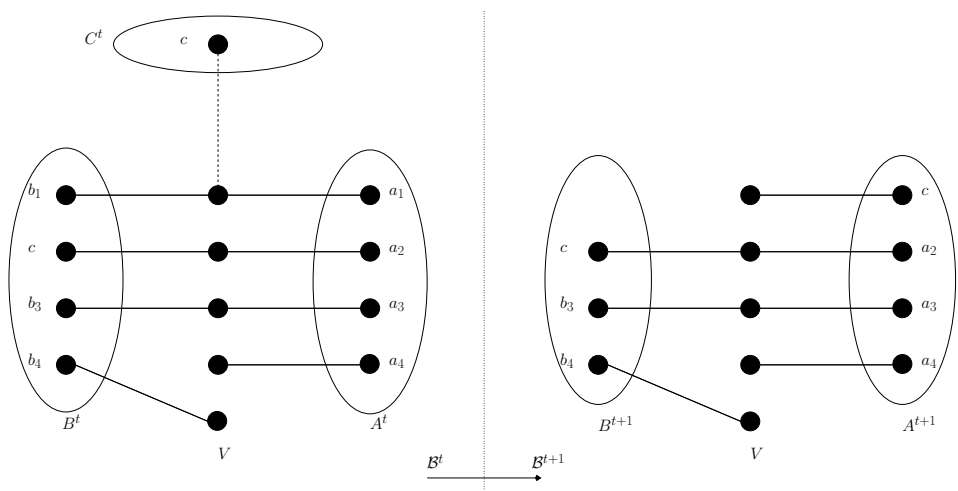

Illustration is given in Figure~\ref{fig:gammaUpdate}.$\hfill\square$

\section{\label{sec:Omitted-proofs}Omitted proofs}

\textbf{Lemma~\ref{lem:math-programming-bound}.} \emph{Let $OPT$
be the optimal feasible strategy for the stochastic probing problem
in our general setting, then $\ex{f\br{OPT}}\le f^{+}\br{x^{+}\cdot p}$.}\\
\\
\emph{Proof.} We construct a feasible solution $x$ of the following
program
\[
\text{maximize }\setst{f^{+}\br{x\cdot p}}{x\cdot p\in\cap_{j=1}^{\kin}\P{\M_{j}^{in}},x\in\cap_{j=1}^{\kout}\P{\M_{j}^{out}}},
\]
by setting $x_{e}=\pr{OPT\mbox{ probes }e}$. First, we show that
this is indeed a feasible solution. Since $OPT$ is a feasible strategy,
the set of elements $Q$ probed by any execution of $OPT$ is always
an independent set of each outer matroid ${\cal M}_{j}^{out}=\br{E,\Iout_{j}}$,
i.e.\ $Q\in\bigcap_{j=1}^{\kout}\Iout_{j}$. Thus the vector $\ex{\chr Q}=x$
may be represented as a convex combination of vectors from $\setst{\chr A}{A\in\bigcap_{j=1}^{\kout}\Iout_{j}}$,
and so $x\in{\cal P}\br{\M_{j}^{out}}$ for any $j\in\set{1,\ldots,\kout}$.
Analogously, the set of elements $S$ that were successfully probed
by $OPT$ satisfy $S\in\bigcap_{j=1}^{\kin}\Iin_{j}$ for every possible
execution of $OPT$. Hence, the vector $\ex{\chr S}=x\cdot p$ may
be represented as a convex combination of vectors from $\setst{\chr A}{A\in\bigcap_{j=1}^{\kin}\Iin_{j}}$
and so $x\cdot p\in{\cal P}\br{\Min_{j}^{in}}$ for any $j\in\set{1,\ldots,\kin}$.
The value $f^{+}(x\cdot p)$ gives the maximum value
of $\exls{S\sim\mathcal{D}}{f(S)}$ over all distributions $\mathcal{D}$
satisfying $\prls{S\sim\mathcal{D}}{e\in S}\leq x_{e}p_{e}$. The
solution $S$ returned by $OPT$ satisfies $\pr{e\in S}=\pr{\text{OPT probes \ensuremath{e}}}p_{e}=x_{e}p_{e}$.
Thus, $OPT$ defines one such distribution, and so we have $\ex{f(OPT)}\le f^{+}(x\cdot p)\le f^{+}(x^{+}\cdot p)$.
$\hfill\square$\\
\\
\textbf{Lemma~\ref{lem:blockingset}.}\emph{ If $p\cdot x\in{\cal P}\br{\M^{in}}$,
then for any element $e$, it holds that\newline $\exls{{\cal C},R\br x}{\sum_{f\in\Gamma^{0}\br e}p_{f}\cdot\indi{f\in R\br x}}\leq1$,
where the expectation is over $R\br x$ and the choice of critical
sets ${\cal C}$; here $\indi{{\cal E}}$ is a 0-1 indicator of random
event ${\cal E}$.}\\
\\
\emph{Proof.} In what follows let us skip writing
$0$ in the superscript of bases $B_{i}^{0}$, mappings $\phi^{0}$,
and set $\Gamma^{0}\br e$. 

Let us condition for now on the critical set $B_{c\br e}$ of element
$e$. For $f$ to belong to $\Gamma\br e$ it has to be the case that
$\phi\brq{B_{c\br f},B_{c\br e}}\br f=e.$ Therefore 
\[
\sum_{f\in\Gamma\br e}p_{f}\cdot\indi{f\in R\br x}=\sum_{f\in E\setminus\set e}p_{f}\cdot\indi{f\in R\br x}\cdot\br{\sum_{i:\phi\brq{B_{i},B_{c\br e}}\br f=e}\indi{B_{i}\mbox{ is }f\mbox{-critical}}},
\]
and by changing the order of summation it is equal to 
\[
\sum_{i}\sum_{f\in B_{i}\setminus e:\phi\brq{B_{i},B_{c\br e}}\br f=e}p_{f}\cdot\indi{f\in R\br x}\cdot\indi{B_{i}\mbox{ is }f\mbox{-critical}}.
\]
Consider $f$ such that $f\in B_{i}\setminus e:\phi\brq{B_{i},B_{c\br e}}\br f=e$.
Since $\indi{f\in R\br x}$ and $c\br f$ (the index critical set
of $f$) are independent, and $\ex{\indi{f\in R\br x}}=x_{f}$ and
$\prcond{B_{i}\mbox{ is }f\mbox{-critical}}{B_{c\br e}}=\prcond{i=c\br f}{B_{c\br e}}=\frac{\beta_{i}}{p_{f}\cdot x_{f}}$,
we get that 
\[
\excondls{}{p_{f}\cdot\indi{f\in R\br x}\cdot\indi{B_{i}^{j}\mbox{ is }f\mbox{-critical}}}{B_{c\br e}}=p_{f}x_{f}\cdot\frac{\beta_{i}}{p_{f}x_{f}}=\beta_{i},
\]
and hence 
\begin{multline*}
\excondls{}{\sum_{f\in\Gamma\br e}p_{f}\cdot\indi{f\in R\br x}}{B_{c\br e}}=\sum_{i}\sum_{f\in B_{i}\setminus e:\phi\brq{B_{i},B_{c\br e}}\br f=e}\beta_{i}^{j}\leq\sum_{i}\beta_{i}^{j}=1,
\end{multline*}
where the inequality follows from the fact that for each $B_{i}$
there can be at most one element $f\in B_{i}$ such that $\phi\brq{B_{i},B_{c\br e}}\br f=e$.\textbf{$\hfill\square$}\\ 
\\
\textbf{Lemma~\ref{lem:combining}}
\emph{Consider a probing problem $\br{E,p,\Iin,\Iout}$.
Suppose we have a $\br{b,c_{out}}$-balanced CR-scheme $\pi^{out}$
for $\P{\Iout}$, and a $\br{b,c_{in}}$-balanced ordered
CR scheme $\pi^{in}$ for $\P{\Iin}$. Then there exists a $\br{b,c_{out}\cdot c_{in}}$-balanced
stoch-CR scheme for $\P{\Iin,\Iout}$.}\\
\\
\emph{Proof.}
First let us recall a Lemma from~\cite{DBLP:conf/stoc/VondrakCZ11}.\\
\textbf{Lemma}
[1.6 from~\cite{DBLP:conf/stoc/VondrakCZ11}]
\emph{ Let ${\cal I} = \bigcap_{i} {\cal I}_i $ and
${\cal P}_{\cal I} = \bigcap_{i} {\cal P}_i$.
Suppose each ${\cal P}_{{\cal I}_i}$ has a monotone $(b,c_i)$-balanced CR scheme. 
Then ${\cal P}_{\cal I}$ has a monotone $(b,\prod_{i} c_i)$-balanced CR scheme defined as $\pi_{x}(A)=\bigcap_{i} \pi_{x}^i(A)$ for $A\subseteq N, x\in b {\cal P}_{\cal I}$.}

Suppose we have a CR scheme $\pi^{out}$ for $\mathcal{P}(\mathcal{I}^{out})$ and 
an ordered CR scheme $\pi^{in}$ for $\mathcal{P}(\mathcal{I}^{in})$.
We would like to define the stochastic contention resolution scheme for $\P{\Iin,\Iout}$ just as in the Lemma above, i.e., as
$\pi_{x}(A) = \pi^{out}_{x}(A) \cap \pi^{in}_{p\cdot x}(\act{A})$.
However, we cannot just simply run $\pi^{out}_x$ on $A$, and then $\pi^{in}_{p\cdot x}$ on $A$ again, and take the intersection, because that does not constitute a feasible probing strategy.
Once again, we need to make use of simulated probes to get a stoch-CR scheme that will have a probability
distribution of $\pi^{out}_{x}(A) \cap \pi^{in}_{p\cdot x}(\act{A})$.
How to implement such a strategy?
We first run $\pi^{out}_{x}(A)$ on the set $A$.
Later we use $\pi^{in}_{p\cdot x}$ to scan elements of $A$ in the order
given from the definition of an ordered scheme.
If $\pi_x^{in}$ considers element $e$ such that $e \in A \setminus \pi^{out}_x(A)$,
then we simulate the probe of $e$; if the $e\in \pi^{out}_x(A)$, then we just probe it.
Therefore, the CR-scheme $\pi^{in}_{p\cdot x}$ works in fact on the set $\act{\pi^{out}_x(A) + (A \setminus \pi^{out}_x(A))^{virt}}$, where $(A \setminus \pi^{out}_x(A))^{virt}$ represents simulated probes of elements in $A \setminus \pi^{out}_x(A)$.
Assuming $e\in \pi^{out}_x(A)$, it is easy to see that elements in $\act{A}$ and elements in $\act{\pi^{out}_x(A) + (A \setminus \pi^{out}_x(A))^{virt}}$ have the same probability distribution.
Therefore, 
\begin{multline}
\pr{e\in \pi^{in}_{p\cdot x}(\act{A}) } =  \\
\prcond{e\in \pi^{in}_{p\cdot x}(\act{\pi^{out}_x(A) \cup (A \setminus \pi^{out}_{x}(A))^{virt}}) }{ \pi^{out}_x(A), e\in \pi^{out}_x(A)}.\label{eq:pinpx}
\end{multline}
And the RHS corresponds to a second phase of a feasible probing strategy. 
Thus we have:
\begin{align*}
&\pr{e\in \pi_x(A) }
\\=\ &\pr{e\in  \pi^{out}_x(A) \cap \pi^{in}_{p\cdot x}( \pi^{out}_x(A) \cup (A \setminus \pi^{out}_x(A))^{virt}	)  } \\
=\ &\exls{ \pi^{out}_x}{\indi{e\in  \pi^{out}_x(A) }  \cdot \excondls{\pi^{in}_{p\cdot x}}{\indi{ e\in \pi^{in}_{p\cdot x}(\act{ \pi^{out}_x(A) \cup (A \setminus \pi^{out}_x(A))^{virt}	) }}}{\pi^{out}_x(A), e\in \pi^{out}_x(A)}}.
\end{align*}
Now just note that 
\begin{multline}
\excondls{\pi^{in}_{p\cdot x}}{\indi{ e\in \pi^{in}_{p\cdot x}( \pi^{out}_x(A) \cup (A \setminus \pi^{out}_x(A))^{virt}	) }}{\pi^{out}_x(A), e\in \pi^{out}_x(A)} =\\
\prcond{e\in \pi^{in}_{p\cdot x}(\act{\pi^{out}_x(A) \cup (A \setminus \pi^{out}_x(A))^{virt}}) }{ \pi^{out}_x(A), e\in \pi^{out}_x(A)}=\\
\pr{e\in \pi^{in}_{p\cdot x}(\act{A})},
\end{multline}
from line~\eqref{eq:pinpx}, and so we can simplify the previous expression to
\begin{align*}
\pr{e\in \pi_x(A) }=\ &\exls{ \pi^{out}_x}{\indi{e\in  \pi^{out}_x(A) } \cdot \pr{e\in \pi^{in}_{p\cdot x}(\act{A}) }}\\\
=\ &\exls{ \pi^{out}_x}{\indi{e\in  \pi^{out}_x(A) }} \cdot \pr{e\in \pi^{in}_{p\cdot x}(\act{A}) }\\
=\ &\pr{e\in  \pi^{out}_x(A) } \cdot \pr{e\in \pi^{in}_{p\cdot x}(\act{A}) }.
\end{align*}
Now the analysis just follows the lines of Lemma~1.6 from~\cite{DBLP:conf/stoc/VondrakCZ11}. 
%=\ &\pr{e\in  \pi^{out}_x(A) } \cdot \prcond{ e \in \pi^{in}_x(A) }{ e\in\pi^{out}_x(A)}.\\
We plug $R(x)$ for $A$, and apply expectation on $\R$ conditioned on $e\in \R$ to get:
\begin{align*}
\exls{\R}{\pr{e\in \pi_x(\R) } | e\in \R}= \exls{\R}{\pr{e\in  \pi^{out}_x(\R) } \cdot \pr{e\in \pi^{in}_{p\cdot x}(\act{\R}) }| e\in \R}.
\end{align*}
From the fact that both $\pi^{out},\pi^{in}$ are monotone, and $\act\R$ is an increasing function of $\R$ we get that $\pr{e\in  \pi^{out}_x(\R) }$ and also $ \pr{e\in \pi^{in}_{p\cdot x}(\act{\R}) }$ are increasing functions of $\R$. Thus applying FKG inequality gives us that
\begin{align*}
&\exls{\R}{\pr{e\in \pi_x(\R) }| e\in \R}\\
= &\exls{\R}{\pr{e\in  \pi^{out}_x(\R) } \cdot \pr{e\in \pi^{in}_{p\cdot x}(\act{\R}) }| e\in \R}\\
\geq &\exls{\R}{\pr{e\in  \pi^{out}_x(\R) }| e\in \R} \cdot \exls{\R}{\pr{e\in \pi^{in}_{p\cdot x}(\act{\R}) }| e\in \R}\\
\geq &b\cdot c_{out} \cdot \exls{\R}{\pr{e\in \pi^{in}_{p\cdot x}(\act{\R}) }| e\in \R}.
\end{align*}
Now applying also expectation on $\act\R$, we get finally
\begin{multline}
\exls{\R,\act\R}{\pr{e\in \pi_x(\R) }| e\in \R}\\
\geq  c_{out} \cdot \exls{\R,\act \R}{\pr{e\in \pi^{in}_{p\cdot x}(\act{\R}) }| e\in \R}
\geq p_e\cdot c_{out}\cdot c_{in}.
\end{multline}

Also, directly from equation  $ \pr{e\in \pi_x(A) } = \pr{e\in  \pi^{out}_x(A) } \cdot \pr{e\in \pi^{in}_{p\cdot x}(\act{A}) }$ we get the monotonicity of the stoch CR-scheme $\pi_x$, since both $\pi^{out}_x$ and $\pi^{in}_{p\cdot x}\br {\act \cdot}$ are monotone.
\textbf{$\hfill\square$}\\ 

\section{\label{sec:Martingale-Theory}Martingale Theory}

\begin{definition} Let $\left(\Omega,{\cal F,\mathbb{P}}\right)$
be a probability space, where $\Omega$ is a sample space, ${\cal F}$
is a $\sigma$-algebra on $\Omega$, and $\mathbb{P}$ is a probability
measure on $(\Omega,{\cal F)}$. Sequence $\left\{ {\cal F}_{t}:t\geq0\right\} $
is called a \emph{filtration} if it is an increasing family of sub-$\sigma$-algebras
of ${\cal F}$: ${\cal F}_{0}\subseteq{\cal F}_{1}\subseteq\ldots\subseteq{\cal F}$.
\end{definition} Intuitively speaking, when considering a stochastic
process, $\sigma$-algebra ${\cal F}_{t}$ represents all information
available to us right after making step $t$. In our case $\sigma$-algebra
${\cal F}_{t}$ contains all information about each randomly chosen
element to probe, about outcome of each probe, and about each support
update for every matroid, that happened before or at step $t$. \begin{definition}
A process $\left(Z_{t}\right)_{t\geq0}$ is called a \emph{martingale}
if for every $t\geq0$ all following conditions hold:
\begin{enumerate}
\item random variable $Z_{t}$ is ${\cal F}_{t}$-measurable,
\item $\ex{\left|Z_{t}\right|}<\infty$,
\item $\excond{Z_{t+1}}{\mathcal{F}_{t}}=X_{t}$. 
\end{enumerate}
\end{definition}

\begin{definition} Random variable $\tau:\Omega\mapsto\left\{ 0,1,\ldots\right\} $
is called a \emph{stopping time} if $\left\{ \tau=t\right\} \in\mathcal{F}_{t}$
for every $t\geq0$. \end{definition} Intuitively, $\tau$ represents
a moment when an event happens. We have to be able to say whether
it happened at step $t$ given only the information from steps $0,1,2,\ldots,t$.
In our case we define $\tau$ as the moment when element became unavailable,
i.e., it was chosen to be probed or it was blocked by other elements.
It is clear that this is a stopping time according to the above definition.\\
\begin{theorem} [Doob's Optional-Stopping Theorem] Let $\left(Z_{t}\right)_{t\geq0}$
be a martingale. Let $\tau$ be a stopping time such that $\tau$
has finite expectation, i.e., $\mathbb{E}[\tau]<\infty$, and the
conditional expectations of the absolute value of the martingale increments
are bounded, i.e., there exists a constant $c$ such that $\mathbb{E}\bigl[|Z_{t+1}-Z_{t}|\,\big\vert\,\mathcal{F}_{t}\bigr]\le c$
for all $t\geq0$. If so, then $\ex{Z_{\tau}}=\ex{Z_{0}}$. \end{theorem}In
our case $\ex{\tau}\leq|A|$, because we just pick an element at random
from $A$, so the expected value of picking $e$ to be probed is exactly
$|A|$, and since it can be earlier blocked by other elements, we
have $\ex{\tau}\leq|A|$. Also the martingale we use is $\br{\br{1+\sum_{f\in\Gamma^{t}\br e}p_{f}X_{f}}\cdot P_{e}^{t}+Y_{e}^{t}}_{t\geq0}$,
and since $P_{e}^{t},Y_{e}^{t}\in\{0,1\}$ then it means that for
any $t$ we have$\size{\br{1+\sum_{f\in\Gamma^{t}\br e}p_{f}X_{f}}\cdot P_{e}^{t}+Y_{e}^{t}}\leq|A|+1$,
and so $\ex{\size{Z_{t+1}-Z_{t}}{\cal F}_{t}}\leq|A|+1$. Therefore,
we can use Doob's Optional-Stopping Theorem in our analysis.

\section{\label{sec:Full-proof-of}Full proof of Lemma~\ref{lem:stochCRscheme}}

\begin{algorithm}
\caption{\label{alg:fullscheme}Stoch-CR scheme $\bar{\pi}_{x}\protect\br A$
for $\protect\M_{1}^{in},...,\protect\M_{\protect\kin}^{in}$ inner
matroids and $\protect\M_{1}^{out},...,\protect\M_{\protect\kout}^{out}$
outer matroids, and $x\in b\cdot\protect\P{\protect\Iin,\protect\Iout}$.}

\begin{algorithmic}[1]

\STATE //Preprocessing:

\STATE find support $\B{}_{in[j]}^{0}$ of $\frac{1}{b}p\cdot x\in{\cal P}\br{\M_{j}^{in}}$
for each $\M_{j}^{in}$;

\STATE find support $\B{}_{out[j]}^{0}$ of $\frac{1}{b}x\in{\cal P}\br{\M_{j}^{out}}$
for each $\M_{j}^{out}$;

\STATE find family $\phi{}_{in[j]}^{0}$ for every $\M_{j}^{in}$;
find family $\phi{}_{out[j]}^{0}$ for every $\M_{j}^{out}$; 

\STATE choose critical sets ${\cal C}_{j}^{in}$ for each $\M_{j}^{in}$
and ${\cal C}_{j}^{out}$ for each $\M_{j}^{out}$ ;

\STATE //Random selection phase:

\STATE remove from $A$ all $e:$ $x_{e}=0$; mark all $e\in A$
as available; $S\leftarrow\emptyset$

\STATE \textbf{while} there are still available elements in $A$
\textbf{do\label{alglabel:while-1}}

\STATE $\qquad$pick element $e$ uniformly at random from $A$\label{alglabel:pick-element-1}

\STATE $\qquad$\textbf{if }$e$ is available \textbf{then}

\STATE $\qquad$$\qquad$probe $e$

\STATE $\qquad$$\qquad$\textbf{if }probe of $e$ successful \textbf{then }

\STATE $\qquad$$\qquad$$\qquad$$S\leftarrow S\cup\set e$

\STATE $\qquad$$\qquad$$\qquad$\textbf{for }each matroid $\M_{j}^{in}$
\textbf{do}

\STATE $\qquad$$\qquad$$\qquad$$\qquad$\textbf{for }each set
$B_{i}^{t}$ of support $\B_{in[j]}^{t}$ \textbf{do}

\STATE $\qquad$$\qquad$$\qquad$$\qquad$$\qquad$$B_{i}^{t}\leftarrow B_{i}^{t}+e$

\STATE $\qquad$$\qquad$call $e$ unavailable

\STATE $\qquad$\textbf{else} simulate the probe of $e$

\STATE $\qquad$\textbf{if }probe or simulation was successful \textbf{then}

\STATE $\qquad$$\qquad$\textbf{for }each set $B_{i}^{t}$ of support
$\B^{t}$ \textbf{do\label{alglabel:supportupdate-random}}

\STATE $\qquad$$\qquad$$\qquad$$f\leftarrow\setphi{B_{c_{j}^{in}\br e}^{t}}{B_{c_{j}^{in}\br f}^{t}}e$

\STATE $\qquad$$\qquad$$\qquad$\textbf{if} $f\neq e$ \textbf{then
}$B_{c_{j}^{in}\br f}^{t}\leftarrow B_{c_{j}^{in}\br f}^{t}-f$ and
call $f$ unavailable\label{alglabel:remove-element-random}

\STATE $\qquad$\textbf{for }each matroid $\M_{j}^{out}$ \textbf{do}

\STATE $\qquad$$\qquad$\textbf{for }each set $B_{i}^{t}$ of support
$\B{}_{out[j]}^{t}$ \textbf{do\label{alglabel:supportupdate-det}}

\STATE $\qquad$$\qquad$$\qquad$$B_{i}^{t}\leftarrow B_{i}^{t}+e$

\STATE $\qquad$$\qquad$$\qquad$$f\leftarrow\setphi{B_{c_{j}^{out}\br e}^{t}}{B_{c_{j}^{out}\br f}^{t}}e$

\STATE $\qquad$$\qquad$$\qquad$\textbf{if} $f\neq e$ \textbf{then
}$B_{c_{j}^{out}\br f}^{t}\leftarrow B_{c_{j}^{out}\br f}^{t}-f$
and call $f$ unavailable\label{alglabel:remove-element-det}

\STATE $\qquad$compute families $\phi{}_{in[j]}^{t+1}$, $\phi{}_{out[j]}^{t+1}$\\
\STATE $\qquad$$t\leftarrow t+1$

\STATE \textbf{return $S$ }as \textbf{$\pi_{x}\br A$}

\end{algorithmic}
\end{algorithm}

The full scheme is presented on Figure~\ref{alg:fullscheme}. Let
us concisely denote by ${\cal C}$ all the critical sets chosen, i.e.,
${\cal C}=\br{{\cal C}_{j}^{in}}_{j\in[\kin]}\times\br{{\cal C}_{j}^{out}}_{j\in[\kout]}$.
Transversal mappings $\phi\br{\M}^{0}$ are found in exactly the same
manner as in the single matroid version.

There are two main differences with respect to what was presented
in the main body. First, since $p\cdot x\in b\cdot\P{\M_{j}^{in}}$,
then $\frac{1}{b}p\cdot x\in\P{\M_{j}^{in}}$, and the support $\B{}_{in[j]}^{0}$
is found by decomposing $\frac{1}{b}p\cdot x$. Thus when element
$f$ chooses a critical set in matroid $\M_{j}^{in}$ it chooses with
probability $\beta_{i}^{in[j]}/\br{\frac{1}{b}p_{f}x_{f}}=b\cdot\beta_{i}^{in[j]}/\br{p_{f}x_{f}}$.
This results in the following modification in Lemma~\ref{lem:blockingset}.
The proof is completely analogous, so we skip it.

\begin{lemma}

\label{lem:blockingset-inner}If $\frac{1}{b}p\cdot x\in{\cal P}\br{\M_{j}^{in}}$,
then for any element $e$, it holds that\\
 $\exls{{\cal C}_{j}^{in},R\br x}{\sum_{f\in\Gamma{}_{in[j]}^{0}\br e}p_{f}\cdot\indi{f\in R\br x}}\leq b$,
where the expectation is over $R\br x$ and the choice of critical
sets ${\cal C}$.

\end{lemma}Now we also need to deal with outer matroids. Again, the
proof is completely analogous, so we skip it.

\begin{lemma}

\label{lem:blockingset-outer}If $\frac{1}{b}x\in{\cal P}\br{\M_{j}^{out}}$,
then for any element $e$, it holds that\\
 $\exls{{\cal C}_{j}^{out},R\br x}{\sum_{f\in\Gamma{}_{out[j]}^{0}\br e}\indi{f\in R\br x}}\leq b$,
where the expectation is over $R\br x$ and the choice of critical
sets ${\cal C}$.

\end{lemma}

Second. Let $Y_{e}^{t}$ for $t=0,1,...$, be a random variable indicating
if $e$ is still available after step $t$. Initially $Y_{e}^{0}=X_{e}$.
Let $P_{e}^{t}$ be a random variable indicating, if $e$ was probed
in one of steps $0,1,...,\mbox{ or }t$. In step $t+1$ element $e$
can be blocked if, for some $j\in[\kin]$, we pick element $f\in\Gamma{}_{in[j]}^{t}\br e$
and successfully probe it (or successfully simulate), or if we just
pick element $f\in\Gamma{}_{out[j]}^{t}\br e$, and probe $f$ or
simulate its probe, irregardless of the outcome. Let $\Gamma_{out}^{t}\br e=\bigcup_{j\in[\kout]}\Gamma{}_{out[j]}^{t}\br e$
and let $\Gamma_{in}^{t}\br e=\bigcup_{j\in[\kin]}\Gamma{}_{in[j]}^{t}\br e\setminus\Gamma_{out}^{t}\br e$
--- this subtraction is to not count an element twice, because if
element $f$ belongs to both $\Gamma{}_{in[j]}^{t}\br e$ and $\Gamma{}_{out[j]}^{t}\br e$,
then just probing $f$ (or simulating its probe) automatically blocks
$e$, irregardless of $f$'s probe (simulation) outcome. Hence, we
do cannot account for the excessive $p_{f}X_{f}$ influence of $f$
on $e$. Therefore, the probability that $e$ stops to be available
at step $t+1$ is equal to
\[
\excond{Y_{e}^{t}-Y_{e}^{t+1}}{{\cal F}^{t},{\cal C}}=\frac{Y_{e}^{t}}{\size A}+\frac{Y_{e}^{t}}{\size A}\cdot\sum_{f\in\Gamma_{in}^{t}\br e}p_{f}X_{f}+\frac{Y_{e}^{t}}{\size A}\cdot\sum_{f\in\Gamma{}_{out}^{t}\br e}X_{f}.
\]
As in the single matroid case, the probability of probing $e$ is
just equal to $\excond{P_{e}^{t+1}-P_{e}^{t}}{{\cal F}^{t},{\cal C}}=\frac{Y_{e}^{t}}{\size A}$.
Thus the martingale we use now in the analysis is
\[
\br{\br{1+\sum_{f\in\Gamma_{in}^{t}\br e}p_{f}X_{f}+\sum_{f\in\Gamma{}_{out}^{t}\br e}X_{f}}\cdot P_{e}^{t}+Y_{e}^{t}}_{t\geq0}.
\]
Let $\tau=\min\setst t{Y_{e}^{t}=0}$ be the step in which edge $e$
became unavailable. It is clear that $\tau$ is a stopping time. Thus
from Doob's stopping theorem we get that\\
\\
\begin{multline*}
\excondls{\tau}{\br{1+\sum_{f\in\Gamma_{in}^{\tau}\br e}p_{f}X_{f}+\sum_{f\in\Gamma{}_{out}^{\tau}\br e}X_{f}}\cdot P_{e}^{\tau}+Y_{e}^{\tau}}{{\cal C}}\\
=\excondls{}{\br{1+\sum_{f\in\Gamma_{in}^{0}\br e}p_{f}X_{f}+\sum_{f\in\Gamma{}_{out}^{0}\br e}X_{f}}\cdot P_{e}^{0}+Y_{e}^{0}}{{\cal C}},
\end{multline*}
 \\
And as before, since the transversal mappings are controlled per matroid,
we have that $\Gamma_{in}^{t}\br e=\Gamma_{in}^{0}\br e$ and $\Gamma_{out}^{t}\br e=\Gamma_{out}^{0}\br e$
for $t\leq\tau$. Thus 
\begin{equation}
\prcond{e\in\bar{\pi}_{x}\br A}{{\cal C}}=p_{e}\cdot\excondls{\tau}{P_{e}^{\tau}}{{\cal C}}=p_{e}X_{e}\left/\br{1+\sum_{f\in\Gamma_{in}^{0}\br e}p_{f}X_{f}+\sum_{f\in\Gamma{}_{out}^{0}\br e}X_{f}}.\right.\label{eq:duze}
\end{equation}
Monotonicity follows as before. The approximation guarantee similarly
from Jensen.

We take random set $\R$ instead of $A$; now $X_{f}=\indi{f\in R\br x}$
is a random variable. Let us condition on $e\in R\br x$, take expected
value $\excondls{{\cal C},R\br x}{\cdot}{e\in R\br x}$ on both sides
of~\ref{eq:duze}, and apply Jensen's inequality to convex function
$x\mapsto\frac{1}{x}$ to get:\\
 \\
$\prcond{e\in\bar{\pi}_{x}\br{\R}}{e\in R\br x}=\excondls{{\cal C},R\br x}{\prcond{e\in\bar{\pi}_{x}\br{\R}}{{\cal C},R\br x}}{e\in R\br x}=$\\
$=\excondls{{\cal C},R\br x}{p_{e}X_{e}\left/\br{1+\sum_{f\in\Gamma_{in}^{0}\br e}p_{f}X_{f}+\sum_{f\in\Gamma{}_{out}^{0}\br e}X_{f}}\right.}{e\in R\br x}$

$\hfill\geq p_{e}\left/\excondls{{\cal C},R\br x}{1+\sum_{f\in\Gamma_{in}^{0}\br e}p_{f}X_{f}+\sum_{f\in\Gamma{}_{out}^{0}\br e}X_{f}}{e\in R\br x}.\right.$\\
\\
Since 
\[
\excondls{{\cal C},R\br x}{\sum_{f\in\Gamma{}_{in[j]}^{0}\br e}p_{f}X_{f}}{e\in R\br x}\leq b
\]
from Lemma~\ref{lem:blockingset-inner}, and 
\[
\excondls{{\cal C},R\br x}{\sum_{f\in\Gamma{}_{out[j]}^{0}\br e}X_{f}}{e\in R\br x}\leq b
\]
from Lemma~\ref{lem:blockingset-outer}, we conclude that
\begin{align*}
 & \,\excondls{{\cal C},R\br x}{1+\sum_{f\in\Gamma{}_{in}^{0}\br e}p_{f}X_{f}+\sum_{f\in\Gamma{}_{out}^{0}\br e}X_{f}}{e\in R\br x}\\
\leq & \,\excondls{{\cal C},R\br x}{1+\sum_{j\in[\kin]}\sum_{f\in\Gamma{}_{in[j]}^{0}\br e}p_{f}X_{f}+\sum_{j\in[\kout]}\sum_{f\in\Gamma{}_{out[j]}^{0}\br e}X_{f}}{e\in R\br x}\\
\leq & \,1+\sum_{j\in[\kin]}\excondls{{\cal C},R\br x}{\sum_{f\in\Gamma{}_{in[j]}^{0}\br e}p_{f}X_{f}}{e\in R\br x}\\
 & \qquad+\sum_{j\in[\kout]}\excondls{{\cal C},R\br x}{\sum_{f\in\Gamma{}_{out[j]}^{0}\br e}X_{f}}{e\in R\br x}\\
\leq & \,1+\kin\cdot b+\kout\cdot b.
\end{align*}
And therefore 
\[
\prcond{e\in\bar{\pi}_{x}\br{\R}}{e\in R\br x}\geq\frac{p_{e}}{b\br{\kin+\kout}+1},
\]
which yields 
\[
\prcond{e\in\bar{\pi}_{x}\br{\R}}{e\in\act{R\br x}}\geq\frac{1}{b\br{\kin+\kout}+1},
\]
which is exactly Property 3 from the definition of stoch-CR scheme.
Lemma~\ref{lem:stochCRscheme} follows.

\section{\label{sec:Stochastic--set-packing}Stochastic $k$-set packing}

In this section we are showing a $(k+1)$-approximation algorithm for Stochastic $k$-Set Packing.

We are given $n$ elements/columns, where each item $e\in E=\brq n$
has a profit $v_{e}\in\mathbb{R}_{+}$, and a random $d$-dimensional
size $S_{e}\in\{0,1\}^{d}$. The sizes are independent for different
items. Additionally, for each item $e$, there is a set $C_{e}$ of
at most $k$ coordinates such that each size vector $S_{e}$ takes
positive values only in these coordinates, i.e., $S_{e}\subseteq C_{e}$
with probability 1. We are also given a capacity vector $b\in\mathbb{Z}_{+}^{d}$
into which items must be packed. We assume that $v_{e}$ is a random
variable that can be correlated with $S_{e}$. The coordinates of
$S_{e}$ also might be correlated between each other.

Important thing to notice is that in this setting, unlike in the previous
ones, here when we probe an element, there is no success/failure outcome.
The size $S_{e}$ of an element $e$ materializes, and the reward
$v_{e}$ is just drawn.

Let $p_{e}^{j}=\ex{S_{e}\br j}$ be the expected size of the $j$
coordinate of column $e$. The following is an LP that models the
problem. Here $U\br c$ denotes a uniform matroid of rank $c$. 
\begin{eqnarray*}
\max\qquad\sum_{e=1}^{n}\ex{v_{e}}\cdot x_{e} &  & \qquad(\mbox{LP-}k\mbox{-set)}\\
\mbox{s.t.}\qquad p^{j}\cdot x\in\P{U\br{b_{j}}} &  & \qquad\forall j\in\brq d\\
x_{e}\in\brq{0,1} &  & \qquad\forall e\in\brq n.
\end{eqnarray*}
Where, as usual, $x_{e}$ stands for $\pr{OPT\mbox{ probes column }e}$.
We are going to present a probing strategy in which for every element
$e$ probability that we will probe $e$ will be at least $\frac{x_{e}}{k+1}$.
From this the Theorem will follow.

The algorithm is presented on Figure~\ref{alg:Stochastic-contention-resolution-k-set-packing}.

\begin{algorithm}
\caption{\label{alg:Stochastic-contention-resolution-k-set-packing}Algorithm
for stochastic $k$-set packing}

\begin{algorithmic}[1]

\STATE //Preprocessing: 

\STATE \textbf{for} each $j\in[d]$ \textbf{do}

\STATE $\qquad$find support $\B_{j}^{0}$ of $p^{j}\cdot x$ in
$\P{U\br{b_{j}}}$\\
\STATE $\qquad$family $\phi_{j}^{0}$ 

\STATE $\qquad$critical sets ${\cal C}=\br{{\cal C}^{j}}_{j\in[d]}$

\STATE //Rounding:

\STATE let $A\leftarrow R\br x$; mark all $e\in A$ as available;
$S\leftarrow\emptyset$

\STATE \textbf{while} there are still available elements in $A$
\textbf{do\label{alglabel:while-2}}

\STATE $\qquad$pick element $e$ uniformly at random from $A$\label{alglabel:pick-element-2}

\STATE $\qquad$\textbf{if }$e$ is available \textbf{then}

\STATE $\qquad$$\qquad$probe $e$

\STATE $\qquad$$\qquad$$S\leftarrow S+e$

\STATE $\qquad$$\qquad$\textbf{for }each $j\in C_{e}$ such that
$S_{e}\br j=1$ \textbf{do}

\STATE $\qquad$$\qquad$$\qquad$\textbf{for }each set $B_{i}^{j,t}$
of support $\B^{j,t}$ \textbf{do}

\STATE $\qquad$$\qquad$$\qquad$$\qquad$$B_{i}^{j,t}\leftarrow B_{i}^{j,t}+e$

\STATE $\qquad$$\qquad$call $e$ unavailable

\STATE $\qquad$\textbf{else} simulate the probe of $e$

\STATE $\qquad$\textbf{for }each $j\in C_{e}$ such that $S_{e}\br j=1$
(whether we probe or simulate) \textbf{do}

\STATE $\qquad$$\qquad$\textbf{for }each set $B_{i}^{j,t}$ of
support $\B^{j,t}$ \textbf{do\label{alglabel:supportupdate-1}}

\STATE $\qquad$$\qquad$$\qquad$$f\leftarrow\setphi{B_{c^{j}\br e}^{j,t}}{B_{c^{j}\br f}^{j,t}}e$

\STATE $\qquad$$\qquad$$\qquad$\textbf{if} $f\neq e$ \textbf{then
}$B_{c^{j}\br f}^{j,t}\leftarrow B_{c^{j}\br f}^{j,t}-f$ and call
$f$ unavailable\label{alglabel:remove-element-1}

\STATE $\qquad$$\qquad$compute the family $\phi_{j}^{t+1}$

\STATE $\qquad$\textbf{for} each $i$ \textbf{do} $B_{i}^{j,t+1}\leftarrow B_{i}^{j,t}$

\STATE $\qquad$$t\leftarrow t+1$

\STATE \textbf{return $S$}

\end{algorithmic}
\end{algorithm}

Constraint for row $j$ is in fact given be a uniform matroid in which
we can take at most $b_{j}$ elements from subset $\setst e{j\in C_{e}}\subseteq E.$
Therefore, we can decompose $p^{j}\cdot x=\sum_{l}\beta_{l}^{j}\cdot B_{l}^{j,0}$.
Uniform matroid is a transversal matroid, so we use the transversal
mapping $\phi_{j}^{0}$ between sets $B^{j,0}$, also let ${\cal C}=\br{{\cal C}^{j}}_{j\in[d]}$
be the vector indicating the critical sets. We define in the same
way as we did already in Lemma~\ref{lem:stochCRscheme}, the sets
$\Gamma_{j}^{0}\br e$ of blocking elements, i.e., 
\[
\Gamma_{j}^{0}\br e=\setst f{f\neq e\wedge\phi_{j}^{0}[B_{c^{j}\br f}^{0},B_{c^{j}\br e}^{0}]}.
\]

As before, let us from now on condition on ${\cal C}$. Let us analyze
the impact of $f\in\Gamma_{j}^{t}\br e$ on $e$. Element $f\in\Gamma_{j}^{t}\br e$
blocks $e$ when $f$ is chosen and $S_{f}\br j=1$. However, right
now $f$ can belong to $\Gamma_{j}^{t}\br e$ for many $j\in C_{e}$.
Therefore if $f$ is chosen in line~\ref{alglabel:pick-element-2}
of the Algorithm~\ref{alg:Stochastic-contention-resolution-k-set-packing},
then the probability that $f$ blocks $e$ is equal to $\prcond{\bigvee_{j:f\in\Gamma_{j}^{t}\br e}\br{S_{f}\br j=1}}{{\cal C}}$.

Let us now repeat the steps of Lemma~1. Let $X_{e}=\indi{e\in R\br x}$.
Let $Y_{e}^{t}$ for $t=0,1,...$, be a random variable indicating
if $e$ is still available after step $t$. Initially $Y_{e}^{0}=X_{e}$.
Let $P_{e}^{t}$ be a random variable indicating, if $e$ was probed
in one of steps $0,1,...,\mbox{ or }t$; we have $P_{e}^{0}=0$ for
all $e$.

Variable $P_{e}^{t+1}-P_{e}^{t}$ indicates if $e$ was probed at
step $t+1$. Given the information ${\cal F}^{t}$ about the process
up to step $t$, the probability of this event is $\excond{P_{e}^{t+1}-P_{e}^{t}}{{\cal F}^{t},{\cal C}}=\frac{Y_{e}^{t}}{\size A}$,
because if element $e$ is still available after step $t$ (i.e.,
$Y_{e}^{t}=1$), then with probability $\frac{1}{\size A}$ we choose
it in line~\ref{alglabel:pick-element-2}, and otherwise (i.e. $Y_{e}^{t}=0$)
we cannot probe it.

Variable $Y_{e}^{t}-Y_{e}^{t+1}$ indicates whether element $e$ stopped
being available at step $t+1$. For this to happen we need to pick
$f\in\Gamma_{j}^{t}\br 0$ and the probe (or simulation) of $f$ needs
to result in a vector $S_{f}$ such that $S_{f}\br j=1$. However,
as already noted, there can be many $j$ for which $f\in\Gamma_{j}^{t}\br 0$.
Therefore, probability that $e$ stops being available in step $t+1$
is equal to 
\[
\excond{Y_{e}^{t}-Y_{e}^{t+1}}{{\cal F}^{t},{\cal C}}=\frac{Y_{e}^{t}}{\size A}+\frac{Y_{e}^{t}}{\size A}\cdot\sum_{f}X_{f}\cdot\prcond{\bigvee_{j:f\in\Gamma_{j}^{t}\br e}\br{S_{f}\br j=1}}{{\cal C}}.
\]
Here we stress the condition on ${\cal C}$ because sets $\Gamma_{j}^{t}$
are constructed given the choice of critical sets. Again we can reason
that 
\[
\br{\br{1+\sum_{f}X_{f}\cdot\prcond{\bigvee_{j:f\in\Gamma_{j}^{t}\br e}\br{S_{f}\br j=1}}{{\cal C}}}\cdot P_{e}^{t}+Y_{e}^{t}}_{t\geq0}
\]
is a martingale. Let $\tau=\min\setst t{Y_{e}^{t}=0}$ be the step
in which edge $e$ became unavailable. It is clear that $\tau$ is
a stopping time. Thus from Doob's Stopping Theorem we get that\\
\\
\begin{multline*}
\excondls{\tau}{\br{1+\sum_{f}X_{f}\cdot\prcond{\bigvee_{j:f\in\Gamma_{j}^{\tau}\br e}\br{S_{f}\br j=1}}{{\cal C}}}\cdot P_{e}^{\tau}+Y_{e}^{\tau}}{{\cal C}}\\
=\excondls{\tau}{\br{1+\sum_{f}X_{f}\cdot\prcond{\bigvee_{j:f\in\Gamma_{j}^{0}\br e}\br{S_{f}\br j=1}}{{\cal C}}}\cdot P_{e}^{0}+Y_{e}^{0}}{{\cal C}}=X_{e}.
\end{multline*}
We argue again using the properties of transversal mapping $\phi_{j}^{t}$
that we have $\Gamma_{j}^{\tau}\br e=\Gamma_{j}^{0}\br e$ for each
$j\in C_{e}$, since $e$ was available before step $\tau$. And if
so, then 
\[
\excondls{\tau}{\br{1+\sum_{f}X_{f}\cdot\prcond{\bigvee_{j:f\in\Gamma_{j}^{0}\br e}\br{S_{f}\br j=1}}{{\cal C}}}\cdot P_{e}^{\tau}+Y_{e}^{\tau}}{{\cal C}}=X_{e}
\]
and since $\br{1+\sum_{f}X_{f}\cdot\prcond{\bigvee_{j:f\in\Gamma_{j}^{0}\br e}\br{S_{f}\br j=1}}{{\cal C}}}$
is just a number depending on ${\cal C}$, we can say that 
\[
\br{1+\sum_{f}X_{f}\cdot\prcond{\bigvee_{j:f\in\Gamma_{j}^{0}\br e}\br{S_{f}\br j=1}}{{\cal C}}}\cdot\excondls{\tau}{P_{e}^{\tau}}{{\cal C}}=X_{e}.
\]
Note that $\excondls{\tau}{P_{e}^{\tau}}{{\cal C}}=\prcond{e\mbox{ is probed}}{{\cal C}}$,
and conclude that 
\[
\prcond{e\mbox{ is probed}}{{\cal C}}=\Frac{X_{e}}{1+\sum_{f}X_{f}\cdot\prcond{\bigvee_{j:f\in\Gamma_{j}^{0}\br e}\br{S_{f}\br j=1}}{{\cal C}}}.
\]
At this point we reason as follows:

\begin{align*}
 & \sum_{f}X_{f}\cdot\prcond{\bigvee_{j:f\in\Gamma_{j}^{0}\br e}\br{S_{f}\br j=1}}{{\cal C}}\\
\leq & \sum_{f}X_{f}\cdot\sum_{j:f\in\Gamma_{j}^{0}\br e}\prcond{S_{f}\br j=1}{{\cal C}}\\
= & \sum_{f}X_{f}\cdot\sum_{j:f\in\Gamma_{j}^{0}\br e}p_{f}^{j}\\
= & \sum_{j\in C_{e}}\sum_{f\in\Gamma_{j}^{0}\br e}p_{f}^{j}X_{f},
\end{align*}
where inequality just follows simply from the union-bound. Then we
have an identity since event $\br{S_{f}\br j=1}$ is independent of
${\cal C}$ and its probability is just equal to $p_{f}^{j}$. Later
we just change the order of summation. Therefore we have shown that
\[
\prcond{e\mbox{ is probed}}{{\cal C}}\geq\Frac{X_{e}}{1+\sum_{j\in C_{e}}\sum_{f\in\Gamma_{j}^{0}\br e}p_{f}^{j}X_{f}}.
\]
We apply expectation $\excondls{{\cal C},R\br x}{\cdot}{e\in R\br x}$
to both sides, use Jensen's inequality to get
\[
\prcond{e\mbox{ is probed}}{e\in R\br x}\geq\Frac 1{1+\excondls{{\cal C},R\br x}{\sum_{j\in C_{e}}\sum_{f\in\Gamma_{j}^{0}\br e}p_{f}^{j}X_{f}}{e\in R\br x}}.
\]
Now we can say that 
\begin{multline*}
\excondls{{\cal C},R\br x}{\sum_{j\in C_{e}}\sum_{f\in\Gamma_{j}^{0}\br e}p_{f}^{j}X_{f}}{e\in R\br x}\\
=\sum_{j\in C_{e}}\excondls{{\cal C},R\br x}{\sum_{f\in\Gamma_{j}^{0}\br e}p_{f}^{j}X_{f}}{e\in R\br x}\leq\sum_{j\in C_{e}}1=|C_{e}|\leq k,
\end{multline*}
where the last inequality $\excondls{{\cal C},R\br x}{\sum_{f\in\Gamma_{j}^{0}\br e}p_{f}^{j}X_{f}}{e\in R\br x}\leq1$
for each $j$ comes from Lemma~\ref{lem:blockingset}. Hence 
\[
\prcond{e\mbox{ is probed}}{e\in R\br x}\geq\frac{1}{1+k},
\]
which gives 
\[
\pr{e\mbox{ is probed}}=\pr{e\in\R}\cdot\prcond{e\mbox{ is probed}}{e\in R\br x}\geq\frac{x_{e}}{1+k}
\]
as desired.

\section{\label{sec:Stochastic-Matching}  Stochastic Matching and handling
negative correlation}

In the Stochastic Matching problem we are given an undirected graph
$G=(V,E)$. Each edge $e\in E$ is assigned a probability $p_{e}\in(0,1]$
and a weight $w_{e}>0$, and each node $v\in V$ is assigned a \emph{patience}
$t_{v}\in\mathbb{N}^{+}$. Each time an edge is probed and it turns
out to be \emph{present} with probability $p_{e}$, in which case
it is (irrevocably) included in the matching we gradually construct
and it gives profit $w_{e}$. Therefore the inner constraints are
given by intersection of two partition matroids. We can probe at most
$t_{u}$ edges that are incident to node $u$ --- these are outer
constraints which also can be described by intersection of two partition
matroids. Our goal is to maximize the expected weight of the constructed
matching.

Let us consider the bipartite case where $V=\br{A\cup B}$ and $E\subseteq A\times B$.
In this case Bansal et al.~\cite{Bansal:woes} provided an LP-based
$3$-approximation. We shall also obtain this approximation factor.
Here we present a variant of our scheme from Lemma~\ref{lem:stochCRscheme}
that does not use transversal mappings, because in this case they
are trivial. Moreover, previously we required the input of the stoch-CR
scheme to be a set of elements sampled independently, i.e., $\R$.
Now we shall apply the ideas from Lemma~\ref{lem:stochCRscheme},
but we use it on a set of edges that are negatively correlated. Such
a set is returned by the algorithm of Gandhi et al.~\cite{DBLP:journals/jacm/GandhiKPS06}.
Also, since the objective is linear, and not submodular, we do not
have to take care of the monotonicity of the scheme, and therefore
we do not draw edges with repetitions. In fact we just scan rounded
edges according to a random permutation, and therefore the below algorithm
is the same that was presented by Bansal et al.~\cite{Bansal:woes}.
Thus what follows is an alternative analysis of the algorithm from~\cite{Bansal:woes}\@.

Consider the following LP for stochastic matching problem:

\begin{align}
\max & \sum_{e}w_{e}p_{e}x_{e}\label{eq:oldLP}\\
 & \sum_{e\in\delta(v)}p_{e}x_{e}\leq1 & \forall v\in V\label{eq:lp1eq1}\\
 & \sum_{e\in\delta(v)}x_{e}\leq t_{v} & \forall v\in V\label{eq:lp1eq2}\\
 & 0\leq x_{e}\leq1 & \forall e\in E.
\end{align}
Suppose that $\br{x_{e}}_{e\in E}$ is the optimal solution to this
LP. We round the solution with dependent rounding of Gandhi et al.~\cite{DBLP:journals/jacm/GandhiKPS06};
we call the algorithm GKPS. Let $\br{\X e}_{e\in E}$ be the rounded
solution, and denote $\E=\setst{e\in E}{\X e=1}$. From the definition
of dependent rounding we know that:
\begin{enumerate}
\item (Marginal distribution) $\pr{\X e=1}=x_{e}$;
\item (Degree preservation) For any $v\in V$ it holds that 
\[
\sum_{e\in\adj v}\X e\leq\left\lceil \sum_{e\in\adj v}x_{e}\right\rceil \leq t_{v};
\]
\item (Negative correlation) For any $v\in V$ and any subset $S\subseteq\adj v$
of edges incident to $v$ it holds that:
\[
\forall_{b\in\set{0,1}}\pr{\bigwedge_{e\in S}\br{\X e=b}}\leq\prod_{e\in S}\pr{\X e=b}.
\]
\end{enumerate}
Negative correlation property and constraint~\eqref{eq:lp1eq1} imply
that
\begin{equation}
\excond{\sum_{f\in\adj e}p_{f}\X f}{e\in\E}\leq\ex{\sum_{f\in\adj e}p_{f}\X f}=\sum_{f\in\adj e}p_{f}x_{f}\leq2-2p_{e}x_{e}.\label{eq:matching-negative}
\end{equation}
Given the solution $\br{\X e}_{e\in E}$ we execute the selection
algorithm presented on Figure~\ref{alg:Stochastic-matching}. Because
of the Degree preservation property we will not exceed the patience
of any vertex. We say that an edge $e$ is safe if no other edge adjacent
to $e$ was already successfully probed; otherwise edge is \emph{blocked.
}Initially all edges are safe.

\begin{algorithm}
\caption{\label{alg:Stochastic-matching}Algorithm for Stochastic Matching}

\begin{algorithmic}[1]

\STATE Solve the LP; let $x$ an optimal solution;

\STATE let $\hat{X}\in\{0,1\}^{E}$ be a solution rounded using GKPS;
let $\E=\setst e{\hat{X}_{e}=1}$; call every $e\in\E$ \emph{safe}

\STATE \textbf{while} there are still safe elements in $\E$ \textbf{do\label{alglabel:while-2-1}}

\STATE $\qquad$pick element $e$ uniformly at random from safe elements
of $\E$\label{alglabel:pick-element-2-1}

\STATE $\qquad$probe $e$

\STATE $\qquad$\textbf{if} probe successful \textbf{then }

\STATE $\qquad$$\qquad$$S\leftarrow S\cup\{e\}$

\STATE $\qquad$$\qquad$call every $f\in\E\cap\delta\br e$ \emph{blocked}

\STATE \textbf{return $S$}

\end{algorithmic}
\end{algorithm}

The expected outcome of our algorithm is
\begin{multline*}
\sum_{e}\pr{e\mbox{ probed}}\cdot p_{e}\cdot w_{e}=\sum_{e}\pr{\X e=1}\cdot\prcond{e\mbox{ probed}}{\X e=1}\cdot p_{e}\cdot w_{e}\\
=\sum_{e}w_{e}p_{e}x_{e}\cdot\prcond{e\mbox{ probed}}{\X e=1},
\end{multline*}
and we shall show that $\prcond{e\mbox{ probed}}{\X e=1}\geq\frac{1}{3}$
for any edge $e$, which will imply $1/3$-approximation of the algorithm.

From now let us condition that we know the set of edges $\E$ and
we know that $\X e=1$.

Consider a random variable $Y_{e}^{t}$ which indicates if edge $e$
is still in the graph after step $t$. We consider variable $Y_{f}^{t}$
for any edge $f\in E$. Initially we have $Y_{f}^{0}=1$ for any $f\in\E$,
and $Y_{f}^{0}=0$ for $f\notin\E$. Let variable $P_{e}^{t}$ denote
if edge $e$ was probed in one of steps $0,1,...,t$; we have $P_{e}^{0}=0$.

Let $\Sigma^{t}$ be the number of edges that are left after $t$
steps. Variable $P_{e}^{t+1}-P_{e}^{t}$ indicates whether edge $e$
was probed in step $t+1$. Given the information ${\cal F}^{t}$ about
the process up to step $t$, probability of this event is $\excond{P_{e}^{t+1}-P_{e}^{t}}{{\cal F}^{t},\E,\X e=1}=\frac{Y_{e}^{t}}{\Sigma^{t}}$,
i.e., if edge $e$ still exists in the graph after step $t$ (i.e.
$Y_{e}^{t}=1$), then the probability is $\frac{1}{\Sigma^{t}}$,
otherwise it is 0.

Variable $Y_{e}^{t}-Y_{e}^{t+1}$ indicates whether edge $e$ was
blocked from the graph in step $t+1$. Given ${\cal F}^{t}$, probability
of this event is $\excond{Y_{e}^{t}-Y_{e}^{t+1}}{{\cal F}^{t},\E,\X e=1}=\frac{Y_{e}^{t}}{\Sigma^{t}}\cdot\br{\sum_{f\in\adj e}p_{f}Y_{f}^{t}+1}$.

It is immediate to note that $Y_{f}^{t}\leq\X f$ for any edge $f$,
and that $P_{e}^{t+1}-P_{e}^{t}$ is always nonnegative. Hence 
\begin{align*}
 & \excond{\br{\sum_{f\in\adj e}p_{f}\X f+1}\cdot\br{P_{e}^{t+1}-P_{e}^{t}}-\br{Y_{e}^{t}-Y_{e}^{t+1}}}{{\cal F}^{t},\E,\X e=1}\\
\geq & \excond{\br{\sum_{f\in\adj e}p_{f}Y_{f}^{t}+1}\cdot\br{P_{e}^{t+1}-P_{e}^{t}}-\br{Y_{e}^{t}-Y_{e}^{t+1}}}{{\cal F}^{t},\E,\X e=1}=0,
\end{align*}
which means that the sequence 
\[
\br{\br{\sum_{f\in\adj e}p_{f}\X f+1}\cdot P_{e}^{t}-\br{1-Y_{e}^{t}}}_{t\geq0}
\]
is a super-martingale.

Let $\tau=\min\setst t{Y_{e}^{t}=0}$ be the step in which edge $e$
was either blocked or probed. It is clear that $\tau$ is a stopping
time. Thus from Doob's Stopping Theorem --- this time in the variant
for super-martingales, i.e., if $\excond{Z^{t+1}-Z^{t}}{{\cal F}^{t}}\geq0$
, then $\ex{Z^{\tau}}\geq\ex{Z^{0}}$ --- we get that 
\begin{multline*}
\excondls{\tau}{\br{\sum_{f\in\adj e}p_{f}\X f+1}\cdot P_{e}^{\tau}-\br{1-Y_{e}^{\tau}}}{\E,\X e=1}\\
\geq\excond{\br{\sum_{f\in\adj e}p_{f}\X f+1}\cdot P_{e}^{0}-\br{1-Y_{e}^{0}}}{\E,\X e=1},
\end{multline*}
where the expectation above is over the random variable $\tau$ only.
Since $P_{e}^{0}=0$,$Y_{e}^{0}=1$,$Y_{e}^{\tau}=0$ the above inequality
implies that
\[
\excondls{\tau}{\br{\sum_{f\in\adj e}p_{f}\X f+1}\cdot P_{e}^{\tau}}{\E,\X e=1}\geq1.
\]
Since we condition all the time on $\E$ and $\X e=1$ we can write
that
\[
\br{\sum_{f\in\adj e}p_{f}\X f+1}\cdot\excondls{\tau}{P_{e}^{\tau}}{\E,\X e=1}\geq1.
\]
Let us notice that $\excondls{\tau}{P_{e}^{\tau}}{\E,\X e=1}$ is
exactly equal to $\prcond{e\mbox{ probed}}{\E,\X e=1}$. Thus we can
write that 
\[
\prcond{e\mbox{ probed}}{\E,\X e=1}\geq\frac{1}{\sum_{f\in\adj e}p_{f}\X f+1}.
\]
Now we can apply to both sides of the above inequality expectation
over $\E$ but still conditioned on $\X e=1$:
\begin{multline*}
\prcond{e\mbox{ probed}}{\X e=1}=\excondls{\E}{\prcond{e\mbox{ probed}}{\E,\X e=1}}{\X e=1}\\
\geq\excondls{\E}{\frac{1}{\sum_{f\in\adj e}p_{f}\X f+1}}{\X e=1},
\end{multline*}
and from Jensen's inequality, and the fact that $x\mapsto\frac{1}{x}$
is convex, we get that
\[
\excondls{\E}{\frac{1}{\sum_{f\in\adj e}p_{f}\X f+1}}{\X e=1}\geq\frac{1}{\excondls{\E}{\sum_{f\in\adj e}p_{f}\X f+1}{\X e=1}}.
\]
From inequality~\eqref{eq:matching-negative} we get $\excondls{\E}{\sum_{f\in\adj e}p_{f}\X f+1}{\X e=1}\leq\exls{\E}{\sum_{f\in\adj e}p_{f}\X f+1}\leq3-2x_{e}p_{e}\leq3$
and we conclude that 
\[
\prcond{e\mbox{ probed}}{\X e=1}\geq\frac{1}{3}.
\]

\end{document}

%% file: injectionUpdate1.pstex_t
\begin{picture}(0,0)%
\includegraphics{injectionUpdate1.pstex}%
\end{picture}%
\setlength{\unitlength}{4144sp}%
\begingroup\makeatletter\ifx\SetFigFont\undefined%
\gdef\SetFigFont#1#2#3#4#5{%
  \reset@font\fontsize{#1}{#2pt}%
  \fontfamily{#3}\fontseries{#4}\fontshape{#5}%
  \selectfont}%
\fi\endgroup%
\begin{picture}(14333,10824)(3233,-9523)
\put(3601,-5911){\makebox(0,0)[lb]{\smash{{\SetFigFont{29}{34.8}{\rmdefault}{\mddefault}{\updefault}{\color[rgb]{0,0,0}$b_3$}%
}}}}
\put(8101,-8386){\makebox(0,0)[lb]{\smash{{\SetFigFont{29}{34.8}{\rmdefault}{\mddefault}{\updefault}{\color[rgb]{0,0,0}$V$}%
}}}}
\put(7201,389){\makebox(0,0)[lb]{\smash{{\SetFigFont{29}{34.8}{\rmdefault}{\mddefault}{\updefault}{\color[rgb]{0,0,0}$c$}%
}}}}
\put(4951,389){\makebox(0,0)[lb]{\smash{{\SetFigFont{29}{34.8}{\rmdefault}{\mddefault}{\updefault}{\color[rgb]{0,0,0}$C^t$}%
}}}}
\put(3601,-3211){\makebox(0,0)[lb]{\smash{{\SetFigFont{29}{34.8}{\rmdefault}{\mddefault}{\updefault}{\color[rgb]{0,0,0}$b_1$}%
}}}}
\put(12601,-5911){\makebox(0,0)[lb]{\smash{{\SetFigFont{29}{34.8}{\rmdefault}{\mddefault}{\updefault}{\color[rgb]{0,0,0}$b_3$}%
}}}}
\put(17101,-8386){\makebox(0,0)[lb]{\smash{{\SetFigFont{29}{34.8}{\rmdefault}{\mddefault}{\updefault}{\color[rgb]{0,0,0}$V$}%
}}}}
\put(8551,-3211){\makebox(0,0)[lb]{\smash{{\SetFigFont{29}{34.8}{\rmdefault}{\mddefault}{\updefault}{\color[rgb]{0,0,0}$v^{B^t}(b_1)=v^{C^t}(c)$}%
}}}}
\put(8551,-4561){\makebox(0,0)[lb]{\smash{{\SetFigFont{29}{34.8}{\rmdefault}{\mddefault}{\updefault}{\color[rgb]{0,0,0}$v^{B^t}(c)$}%
}}}}
\put(8551,-7261){\makebox(0,0)[lb]{\smash{{\SetFigFont{29}{34.8}{\rmdefault}{\mddefault}{\updefault}{\color[rgb]{0,0,0}$v^{B^t}(b_3)$}%
}}}}
\put(17551,-4561){\makebox(0,0)[lb]{\smash{{\SetFigFont{29}{34.8}{\rmdefault}{\mddefault}{\updefault}{\color[rgb]{0,0,0}$v^{B^{t+1}}(c)$}%
}}}}
\put(17551,-3211){\makebox(0,0)[lb]{\smash{{\SetFigFont{29}{34.8}{\rmdefault}{\mddefault}{\updefault}{\color[rgb]{0,0,0}$v^{C^{t}}(c)$}%
}}}}
\put(10351,-8836){\makebox(0,0)[lb]{\smash{{\SetFigFont{29}{34.8}{\rmdefault}{\mddefault}{\updefault}{\color[rgb]{0,0,0}$\mathcal{B}^t$}%
}}}}
\put(12376,-8836){\makebox(0,0)[lb]{\smash{{\SetFigFont{29}{34.8}{\rmdefault}{\mddefault}{\updefault}{\color[rgb]{0,0,0}$\mathcal{B}^{t+1}$}%
}}}}
\put(17551,-7261){\makebox(0,0)[lb]{\smash{{\SetFigFont{29}{34.8}{\rmdefault}{\mddefault}{\updefault}{\color[rgb]{0,0,0}$v^{B^{t+1}}(b_3)$}%
}}}}
\put(4276,-7486){\makebox(0,0)[lb]{\smash{{\SetFigFont{29}{34.8}{\rmdefault}{\mddefault}{\updefault}{\color[rgb]{0,0,0}$B^t$}%
}}}}
\put(13276,-7486){\makebox(0,0)[lb]{\smash{{\SetFigFont{29}{34.8}{\rmdefault}{\mddefault}{\updefault}{\color[rgb]{0,0,0}$B^{t+1}$}%
}}}}
\put(12601,-4561){\makebox(0,0)[lb]{\smash{{\SetFigFont{29}{34.8}{\rmdefault}{\mddefault}{\updefault}{\color[rgb]{0,0,0}$c$}%
}}}}
\put(3601,-4561){\makebox(0,0)[lb]{\smash{{\SetFigFont{29}{34.8}{\rmdefault}{\mddefault}{\updefault}{\color[rgb]{0,0,0}$c$}%
}}}}
\end{picture}%

%% file: injectionUpdate2.pstex_t
\begin{picture}(0,0)%
\includegraphics{injectionUpdate2.pstex}%
\end{picture}%
\setlength{\unitlength}{4144sp}%
\begingroup\makeatletter\ifx\SetFigFont\undefined%
\gdef\SetFigFont#1#2#3#4#5{%
  \reset@font\fontsize{#1}{#2pt}%
  \fontfamily{#3}\fontseries{#4}\fontshape{#5}%
  \selectfont}%
\fi\endgroup%
\begin{picture}(14333,10824)(3233,-9523)
\put(3601,-5911){\makebox(0,0)[lb]{\smash{{\SetFigFont{29}{34.8}{\rmdefault}{\mddefault}{\updefault}{\color[rgb]{0,0,0}$b_3$}%
}}}}
\put(3601,-4561){\makebox(0,0)[lb]{\smash{{\SetFigFont{29}{34.8}{\rmdefault}{\mddefault}{\updefault}{\color[rgb]{0,0,0}$b_2$}%
}}}}
\put(8101,-8386){\makebox(0,0)[lb]{\smash{{\SetFigFont{29}{34.8}{\rmdefault}{\mddefault}{\updefault}{\color[rgb]{0,0,0}$V$}%
}}}}
\put(7201,389){\makebox(0,0)[lb]{\smash{{\SetFigFont{29}{34.8}{\rmdefault}{\mddefault}{\updefault}{\color[rgb]{0,0,0}$c$}%
}}}}
\put(4951,389){\makebox(0,0)[lb]{\smash{{\SetFigFont{29}{34.8}{\rmdefault}{\mddefault}{\updefault}{\color[rgb]{0,0,0}$C^t$}%
}}}}
\put(3601,-3211){\makebox(0,0)[lb]{\smash{{\SetFigFont{29}{34.8}{\rmdefault}{\mddefault}{\updefault}{\color[rgb]{0,0,0}$b_1$}%
}}}}
\put(12601,-3211){\makebox(0,0)[lb]{\smash{{\SetFigFont{29}{34.8}{\rmdefault}{\mddefault}{\updefault}{\color[rgb]{0,0,0}$c$}%
}}}}
\put(12601,-4561){\makebox(0,0)[lb]{\smash{{\SetFigFont{29}{34.8}{\rmdefault}{\mddefault}{\updefault}{\color[rgb]{0,0,0}$b_2$}%
}}}}
\put(12601,-5911){\makebox(0,0)[lb]{\smash{{\SetFigFont{29}{34.8}{\rmdefault}{\mddefault}{\updefault}{\color[rgb]{0,0,0}$b_3$}%
}}}}
\put(17101,-8386){\makebox(0,0)[lb]{\smash{{\SetFigFont{29}{34.8}{\rmdefault}{\mddefault}{\updefault}{\color[rgb]{0,0,0}$V$}%
}}}}
\put(8551,-3211){\makebox(0,0)[lb]{\smash{{\SetFigFont{29}{34.8}{\rmdefault}{\mddefault}{\updefault}{\color[rgb]{0,0,0}$v^{B^t}(b_1)=v^{C^t}(c)$}%
}}}}
\put(8551,-4561){\makebox(0,0)[lb]{\smash{{\SetFigFont{29}{34.8}{\rmdefault}{\mddefault}{\updefault}{\color[rgb]{0,0,0}$v^{B^t}(b_2)$}%
}}}}
\put(8551,-7261){\makebox(0,0)[lb]{\smash{{\SetFigFont{29}{34.8}{\rmdefault}{\mddefault}{\updefault}{\color[rgb]{0,0,0}$v^{B^t}(b_3)$}%
}}}}
\put(17551,-4561){\makebox(0,0)[lb]{\smash{{\SetFigFont{29}{34.8}{\rmdefault}{\mddefault}{\updefault}{\color[rgb]{0,0,0}$v^{B^{t+1}}(b_2)$}%
}}}}
\put(17551,-3211){\makebox(0,0)[lb]{\smash{{\SetFigFont{29}{34.8}{\rmdefault}{\mddefault}{\updefault}{\color[rgb]{0,0,0}$v^{B^{t+1}}(c)$}%
}}}}
\put(10351,-8836){\makebox(0,0)[lb]{\smash{{\SetFigFont{29}{34.8}{\rmdefault}{\mddefault}{\updefault}{\color[rgb]{0,0,0}$\mathcal{B}^t$}%
}}}}
\put(12376,-8836){\makebox(0,0)[lb]{\smash{{\SetFigFont{29}{34.8}{\rmdefault}{\mddefault}{\updefault}{\color[rgb]{0,0,0}$\mathcal{B}^{t+1}$}%
}}}}
\put(17551,-7261){\makebox(0,0)[lb]{\smash{{\SetFigFont{29}{34.8}{\rmdefault}{\mddefault}{\updefault}{\color[rgb]{0,0,0}$v^{B^{t+1}}(b_3)$}%
}}}}
\put(4276,-7486){\makebox(0,0)[lb]{\smash{{\SetFigFont{29}{34.8}{\rmdefault}{\mddefault}{\updefault}{\color[rgb]{0,0,0}$B^t$}%
}}}}
\put(13276,-7486){\makebox(0,0)[lb]{\smash{{\SetFigFont{29}{34.8}{\rmdefault}{\mddefault}{\updefault}{\color[rgb]{0,0,0}$B^{t+1}$}%
}}}}
\end{picture}%

%% file: gammaUpdate.pstex_t
\begin{picture}(0,0)%
\includegraphics{gammaUpdate.pstex}%
\end{picture}%
\setlength{\unitlength}{4144sp}%
\begingroup\makeatletter\ifx\SetFigFont\undefined%
\gdef\SetFigFont#1#2#3#4#5{%
  \reset@font\fontsize{#1}{#2pt}%
  \fontfamily{#3}\fontseries{#4}\fontshape{#5}%
  \selectfont}%
\fi\endgroup%
\begin{picture}(22426,11634)(3143,-10423)
\put(3601,-5911){\makebox(0,0)[lb]{\smash{{\SetFigFont{29}{34.8}{\rmdefault}{\mddefault}{\updefault}{\color[rgb]{0,0,0}$b_3$}%
}}}}
\put(7201,389){\makebox(0,0)[lb]{\smash{{\SetFigFont{29}{34.8}{\rmdefault}{\mddefault}{\updefault}{\color[rgb]{0,0,0}$c$}%
}}}}
\put(4951,389){\makebox(0,0)[lb]{\smash{{\SetFigFont{29}{34.8}{\rmdefault}{\mddefault}{\updefault}{\color[rgb]{0,0,0}$C^t$}%
}}}}
\put(3601,-3211){\makebox(0,0)[lb]{\smash{{\SetFigFont{29}{34.8}{\rmdefault}{\mddefault}{\updefault}{\color[rgb]{0,0,0}$b_1$}%
}}}}
\put(3601,-4561){\makebox(0,0)[lb]{\smash{{\SetFigFont{29}{34.8}{\rmdefault}{\mddefault}{\updefault}{\color[rgb]{0,0,0}$c$}%
}}}}
\put(12151,-3211){\makebox(0,0)[lb]{\smash{{\SetFigFont{29}{34.8}{\rmdefault}{\mddefault}{\updefault}{\color[rgb]{0,0,0}$a_1$}%
}}}}
\put(12151,-5911){\makebox(0,0)[lb]{\smash{{\SetFigFont{29}{34.8}{\rmdefault}{\mddefault}{\updefault}{\color[rgb]{0,0,0}$a_3$}%
}}}}
\put(12151,-4561){\makebox(0,0)[lb]{\smash{{\SetFigFont{29}{34.8}{\rmdefault}{\mddefault}{\updefault}{\color[rgb]{0,0,0}$a_2$}%
}}}}
\put(16201,-4561){\makebox(0,0)[lb]{\smash{{\SetFigFont{29}{34.8}{\rmdefault}{\mddefault}{\updefault}{\color[rgb]{0,0,0}$c$}%
}}}}
\put(16201,-5911){\makebox(0,0)[lb]{\smash{{\SetFigFont{29}{34.8}{\rmdefault}{\mddefault}{\updefault}{\color[rgb]{0,0,0}$b_3$}%
}}}}
\put(24751,-4561){\makebox(0,0)[lb]{\smash{{\SetFigFont{29}{34.8}{\rmdefault}{\mddefault}{\updefault}{\color[rgb]{0,0,0}$a_2$}%
}}}}
\put(24751,-5911){\makebox(0,0)[lb]{\smash{{\SetFigFont{29}{34.8}{\rmdefault}{\mddefault}{\updefault}{\color[rgb]{0,0,0}$a_3$}%
}}}}
\put(24751,-3211){\makebox(0,0)[lb]{\smash{{\SetFigFont{29}{34.8}{\rmdefault}{\mddefault}{\updefault}{\color[rgb]{0,0,0}$c$}%
}}}}
\put(8101,-9511){\makebox(0,0)[lb]{\smash{{\SetFigFont{29}{34.8}{\rmdefault}{\mddefault}{\updefault}{\color[rgb]{0,0,0}$V$}%
}}}}
\put(11701,-8791){\makebox(0,0)[lb]{\smash{{\SetFigFont{29}{34.8}{\rmdefault}{\mddefault}{\updefault}{\color[rgb]{0,0,0}$A^t$}%
}}}}
\put(3601,-7261){\makebox(0,0)[lb]{\smash{{\SetFigFont{29}{34.8}{\rmdefault}{\mddefault}{\updefault}{\color[rgb]{0,0,0}$b_4$}%
}}}}
\put(12151,-7261){\makebox(0,0)[lb]{\smash{{\SetFigFont{29}{34.8}{\rmdefault}{\mddefault}{\updefault}{\color[rgb]{0,0,0}$a_4$}%
}}}}
\put(4411,-8791){\makebox(0,0)[lb]{\smash{{\SetFigFont{29}{34.8}{\rmdefault}{\mddefault}{\updefault}{\color[rgb]{0,0,0}$B^t$}%
}}}}
\put(13051,-9736){\makebox(0,0)[lb]{\smash{{\SetFigFont{29}{34.8}{\rmdefault}{\mddefault}{\updefault}{\color[rgb]{0,0,0}$\mathcal{B}^t$}%
}}}}
\put(15526,-9736){\makebox(0,0)[lb]{\smash{{\SetFigFont{29}{34.8}{\rmdefault}{\mddefault}{\updefault}{\color[rgb]{0,0,0}$\mathcal{B}^{t+1}$}%
}}}}
\put(16201,-7261){\makebox(0,0)[lb]{\smash{{\SetFigFont{29}{34.8}{\rmdefault}{\mddefault}{\updefault}{\color[rgb]{0,0,0}$b_4$}%
}}}}
\put(24301,-8836){\makebox(0,0)[lb]{\smash{{\SetFigFont{29}{34.8}{\rmdefault}{\mddefault}{\updefault}{\color[rgb]{0,0,0}$A^{t+1}$}%
}}}}
\put(17101,-8836){\makebox(0,0)[lb]{\smash{{\SetFigFont{29}{34.8}{\rmdefault}{\mddefault}{\updefault}{\color[rgb]{0,0,0}$B^{t+1}$}%
}}}}
\put(24751,-7261){\makebox(0,0)[lb]{\smash{{\SetFigFont{29}{34.8}{\rmdefault}{\mddefault}{\updefault}{\color[rgb]{0,0,0}$a_4$}%
}}}}
\put(20701,-9511){\makebox(0,0)[lb]{\smash{{\SetFigFont{29}{34.8}{\rmdefault}{\mddefault}{\updefault}{\color[rgb]{0,0,0}$V$}%
}}}}
\end{picture}%